\documentclass[reprint,amsmath,amssymb,aps,prl]{revtex4-1}
\usepackage[T1]{fontenc}
\usepackage[utf8]{inputenc}
\usepackage{graphicx}
\usepackage{color}
\usepackage{txfonts}
\usepackage{times}
\usepackage{hyperref}
\hypersetup{
    unicode=true,          
    pdftitle={Forbidden clock transitions in weakly bound molecules},    
    pdfauthor={Mateusz Borkowski},     
    colorlinks=true,       
    linkcolor=blue,          
    citecolor=blue,        
    filecolor=blue,      
    urlcolor=blue,           
}

\newcounter{TodoCounter}

\begin{document}
	\title{Optical lattice clocks with weakly bound molecules}
	
	\author{Mateusz Borkowski}
	\email{mateusz@fizyka.umk.pl}
	\affiliation{Institute of Physics, Faculty of Physics, Astronomy and Informatics, Nicolaus Copernicus University, Grudziadzka 5, 87-100 Torun, Poland}

	\date{\today}
	
	\begin{abstract}
		Optical molecular clocks promise unparalleled sensitivity to the temporal variation of the electron-to-proton mass ratio and insight into possible new physics beyond the Standard Model.
		We propose to realize a molecular clock with bosonic $^{174}$Yb$_2$ molecules, where the forbidden $^1$S$_0$$\rightarrow$$^3$P$_0$ clock transition would be induced magnetically.
		The use of a bosonic species avoids possible complications due to hyperfine structure present in fermionic species. 
		While direct clock line photoassociation would be challenging, weakly bound ground state molecules could be produced by STIRAP and used instead. The recent scattering measurements 
		[L. Franchi, \emph{et al.} \href{https://doi.org/10.1088/1367-2630/aa8fb4}{New J. Phys {\bf 19}, 103037 (2017)}] enable us to
		determine the positions of target $^1$S$_0$+$^3$P$_0$ vibrational levels and calculate the Franck-Condon factors for clock transitions between ground and excited molecular states. The resulting magnetically induced Rabi frequencies are similar to those for atoms hinting that an experimental realization is feasible. A successful observation could pave the way towards Hz-level molecular spectroscopy.
\end{abstract}

	\maketitle
	
	Optical atomic clocks \cite{Katori2011, Ludlow2015} have made it possible to test the fundamentals of physics \cite{Chou2010} and place limits on temporal variation of fundamental constants \cite{Lea2007, Fortier2007, Blatt2008, Orzel2012, Ferreira2012, Safronova2017}, to explore  quantum many-body systems \cite{Martin2013, Rey2014,Dareau2015} or even to search for topological dark matter through its impact on the fine-structure constant \cite{Derevianko2014, Wcislo2016}. 
	Molecular clocks promise increased sensitivity to the variation of the electron-to-proton mass ratio~\cite{Shelkovnikov2008, Zelevinsky2008, DeMille2008, Beloy2011}. Observation of coherent oscillations of this fundamental constant could also inform of the presence of dark matter~\cite{Stadnik2015, Arvanitaki2015}. Thanks to their simple structure, molecules composed of alkaline-earth-metal and similar atoms are excellent prototype systems for the investigation of, for example, Casimir-Polder interactions \cite{Zhang2008} or new gravitylike forces~\cite{Salumbides2014, BorkowskiGravityTBP}.	
	For this reason, spectroscopy of molecular or quasi-molecular systems using doubly forbidden transitions is a rapidly growing field: recent achievements include the production of subradiant states in Yb$_2$~\cite{Takasu2012} and Sr$_2$~\cite{McGuyer2015, McGuyer2015a}, magnetic tuning of forbidden molecular $J=0\rightarrow2$ transitions~\cite{McGuyer2015b} and photoassociation near the ultranarrow $^1$S$_0$$\rightarrow$$^3$P$_2$ line~\cite{Taie2016}.
	A molecular clock using the $^1$S$_0$$\rightarrow$$^3$P$_0$ transition, however, has remained elusive.

	\begin{figure}[b]
		\hspace{-1.5mm}\includegraphics[clip, width=0.49\textwidth]{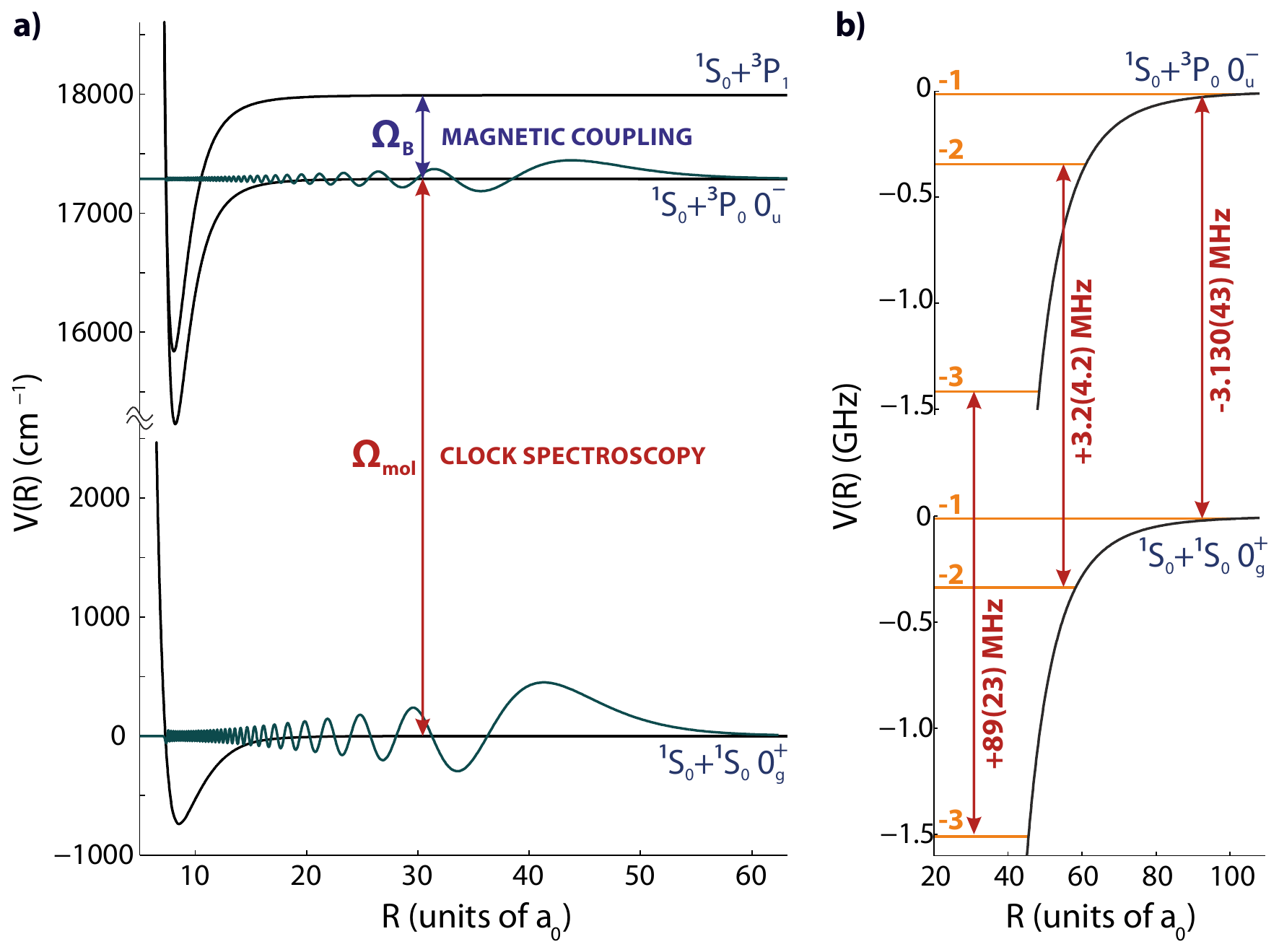}
		\caption{(Color online) 
			a) Magnetic coupling $\Omega_B$ between $^1$S$_0$+$^3$P$_0$ and  $^1$S$_0$+$^3$P$_1$ states (only the $0_u^+$ potential~\cite{Borkowski2009} is shown for clarity) makes it possible to induce an optical Rabi frequency $\Omega_{\rm mol}$ on a molecular clock transition. The radial wavefunctions (in arb. units) for the $v'_g=-3$ and $v'_e=-3$ states are also shown.
			b) Ground- and excited-state bound states and three strongest optical transitions between $^1$S$_0$+$^1$S$_0$ $0_g^+$ and $^1$S$_0$+$^3$P$_0$ $0_u^-$ states. The positions of molecular clock lines are given relative to the atomic clock transition. \label{fig:potentials}}
	\end{figure}	

\begin{figure}[t]
		\hspace{-1.5mm}\includegraphics[clip, width=0.49\textwidth]{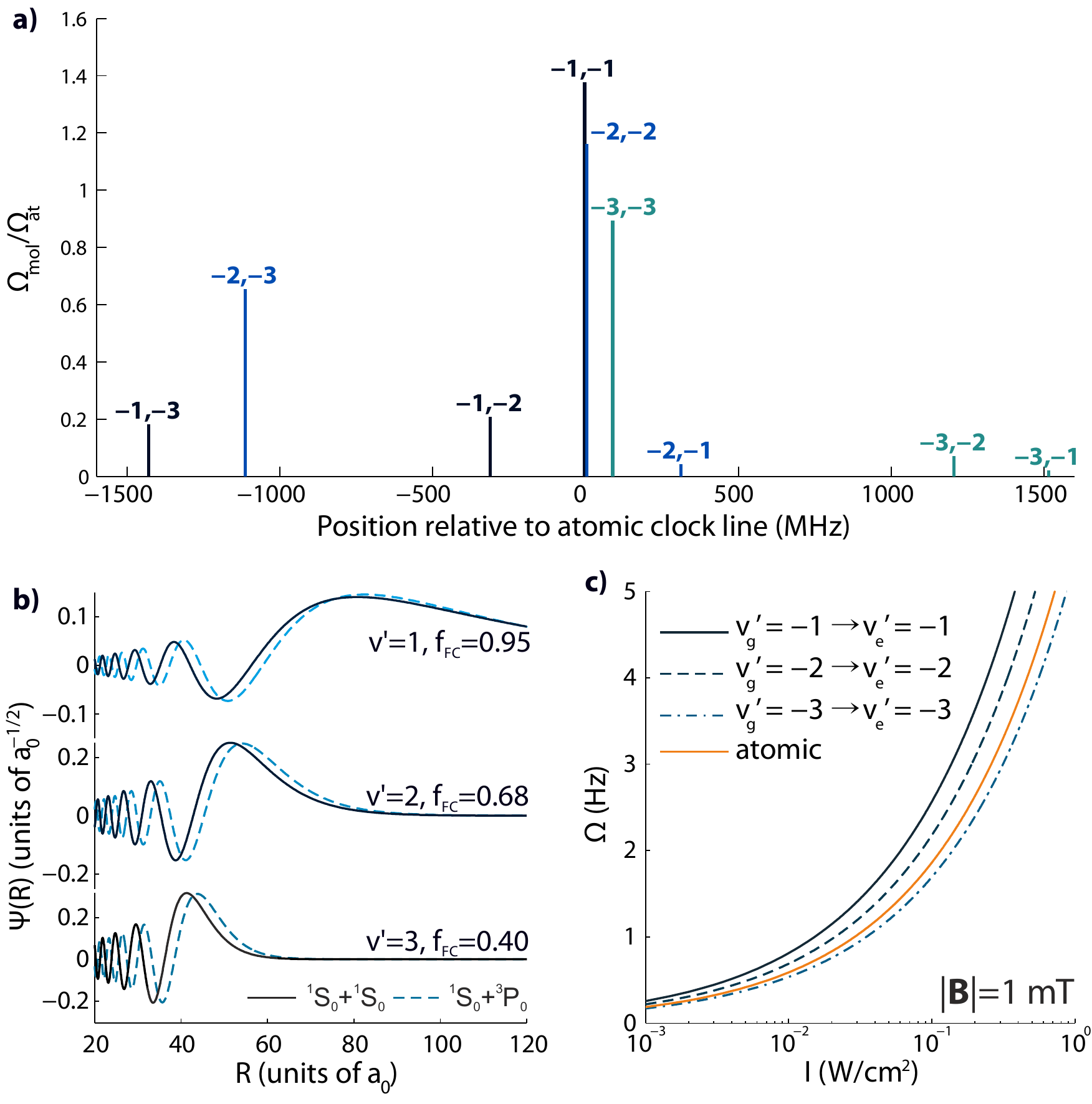}
		\caption{(Color online) a) Positions and relative Rabi frequencies $\Omega_{\rm mol}/\Omega_{\rm at}$ of the molecular clock transitions between ground $^1$S$_0$+$^1$S$_0$ and excited $^1$S$_0$+$^3$P$_0$ states. The numbers indicate the ground and excited vibrational quantum numbers. b) Radial wavefunctions $\Psi(R)$ for the top three energy levels of the rotationless ground $^{174}$Yb$_2$ $^1$S$_0$+$^1$S$_0$ $0_g^+$ and $^1$S$_0$+$^3$P$_0$ $0_u^-$ excited states. 
			c) Magnetically induced optical Rabi frequencies for three transitions with the best Franck-Condon factors for a magnetic field $|\mathbf B|= 1\,{\rm mT}$. 
			\label{fig:boundwf}}
	\end{figure}

	In this Letter we show that clock spectroscopy in a molecular system may indeed be within reach if one chooses to work with weakly bound ultracold molecules. We focus on a spinless bosonic species, $^{174}$Yb$_2$, to avoid possible complications related to the hyperfine structure present in fermions. In bosonic atoms the strictly forbidden $^1$S$_0$-$^3$P$_0$ transition can be enabled by applying a small magnetic field~\cite{Taichenachev2006, Barber2006}. The same mechanism could be used for molecules (Fig.~\ref{fig:potentials}a): clock spectroscopy between molecular $^1$S$_0$+$^1$S$_0$ and $^1$S$_0$+$^3$P$_0$ states would be enabled by magnetic coupling of the latter and the bright $^1$S$_0$+$^3$P$_1$ state. Recent determinations of ground-excited-state~\cite{Franchi2017} and ground-ground-state~\cite{Borkowski2017a} scattering properties allow us to predict the positions of near-threshold $^1$S$_0$+$^3$P$_0$ bound states. Direct observation of these states by clock line photoassociation turns out to be challenging due to unfavorable Franck-Condon factors even in a tight lattice. Instead we propose to produce weakly bound ground state molecules by STIRAP~\cite{Bergmann1998, Stellmer2012, Ciamei2017} using $^1$S$_0$+$^3$P$_1$ intermediate states~\cite{Tojo2006, Borkowski2009} and conduct clock spectroscopy between near-threshold $^1$S$_0$+$^1$S$_0$ and $^1$S$_0$+$^3$P$_0$ bound states (Fig.~\ref{fig:potentials}b). Multiple clock transitions are available, depending on the choice of ground and excited vibrational levels (Fig.~\ref{fig:boundwf}a). Thanks to Franck-Condon factors close to unity the magnetically induced molecular Rabi frequencies can be on the order of or even larger than atomic under similar experimental conditions. In principle, the proposed molecular clock could be implemented in existing $^{174}$Yb atomic clock experiments with 3D optical lattices~\cite{Franchi2017, Bouganne2017}.
	
{\color{blue}}

	\begin{table}[t]
		\caption{Scattering lengths and potential parameters used for the determination of the $^1$S$_0$+$^3$P$_0$ $0_u^-$ energy levels. The accurate measurement of the difference $a_{\rm ge}-a_{\rm gg}$ between ground-excited-state and ground-ground-state $s$-wave scattering lengths \cite{Franchi2017} combined with a recent determination of $a_{\rm gg}$ \cite{Borkowski2017a} gives a precise value of $a_{\rm ge} = +94.84(0.14)\,a_0$. The~$C_6$~coefficient for the $^1$S$_0$+$^3$P$_0$ state was given in Ref.~\cite{Porsev2014}, while the~$C_8$~coefficient was estimated using an asymptotic interaction model \cite{Mies1978} from the $C_8$ coefficients for the $^1$S$_0$+$^3$P$_1$ asymptote. $E_h \approx 4.359745 \times 10^{-18}~\rm{J}$ is the Hartree energy and $a_0 \approx 0.0529177~{\rm nm}$ is the Bohr radius~\cite{Mohr2016}. \label{tbl:parameters}}
		\begin{ruledtabular}
			\begin{tabular}{l r r r l}
				Parameter  & Value &  Unit & Source\\
				\hline
				$\left(a_{\rm ge}-a_{\rm gg}\right)$  & $-10.19(0.13)$ & $a_0$ & Ref. \cite{Franchi2017}\\
				& $-19(11)$ & $a_0$ & Ref. \cite{Bouganne2017} \\
				$a_{\rm gg} $ & $+104.9(1.5)$ & $a_0$& Ref. \cite{Kitagawa2008} \\
				& $+105.0332(95)$ & $a_0$& Ref. \cite{Borkowski2017a}\\
				$a_{\rm ge}$  & $+94.7(1.6)$ & $a_0$& Ref. \cite{Franchi2017}\\
				& $+94.84(0.14)$ & $a_0$ & Refs. \cite{Franchi2017, Borkowski2017a} combined \\
				$C_6(0_u^-)$  & 2561(95) & $E_h a_0^6$ & Ref. \cite{Porsev2014} \\
				$C_8(0_u^+)$  & 3.20(0.14) & $10^5 E_h a_0^8$ &''\mbox{\hspace{5mm}}\\
				$C_8(1_u)$  & 4.11(0.18) & $10^5 E_h a_0^8$ & ''\mbox{\hspace{5mm}} \\
				$C_8(0_u^-)$  & 3.81(0.21) & $10^5 E_h a_0^8$ & Estimated~\footnote{Ref.~\cite{Porsev2014} did not give the $C_8(0_u^-)$ coefficient, but it can be estimated from $C_8(0_u^+)$ and $C_8(1_u)$ using the asymptotic relations between Hund's case (c) $0_u^+$, $1_u$ and $0_u^-$ potentials and the $^1$S+$^3$P case (a) potentials $^3\Sigma_u^+$ and $^3\Pi_u$~\cite{Mies1978}. A~similar estimate for $C_6(0_u^-)$ gives a value 6.9\% larger than Ref.~\cite{Porsev2014} so we added this percentage to the uncertainty evaluation for $C_8(0_u^-)$ as a measure of an additional systematic error.
				 }\\
			\end{tabular}
		\end{ruledtabular}

	\end{table}	

	We start with the determination of $^1$S$_0$+$^3$P$_0$ bound state energies using a model potential. The relative motion of atoms in the $^1$S$_0$+$^3$P$_0$ state can be governed by a $0_g^-$ (\emph{gerade}) or $0_u^-$ (\emph{ungerade}) potential. If the two atoms were originally in the ground $^1$S$_0$ state and were excited close together as a pair, like in the experiments~\cite{Bouganne2017, Franchi2017}, then Laporte rule~\cite{Atkins2010} dictates that the excited pair ends up in the \emph{ungerade} $0_u^-$ excited state. The experimental ground-excited-state scattering length $a_{\rm ge}$ therefore applies to the $0_u^-$ potential. The vibrational spacings of energy levels close to the dissociation limit depend primarily on the van der Waals $-C_6(0_u^-)\,R^{-6}$ interaction \cite{Leroy1970}. On the other hand, the position of the topmost vibrational state is related to the scattering length $a_{\rm ge}$, because both are determined by a zero-energy WKB phase integral of the interaction potential \cite{Gribakin1993, Ciurylo2004, Kitagawa2008}. This relationship is routinely used to determine the $s$-wave scattering lengths and long range interactions via photoassociation spectroscopy of near threshold energy levels~\cite{Jones2006, Lett1993, Abraham1995, Tsai1997, MartinezDeEscobar2008, Kitagawa2008, Munchow2011, Kahmann2014, Borkowski2017a}. 
	
	Here we already have the experimental $^1$S$_0$+$^3$P$_0$ $s$-wave scattering length $a_{ge}$ from the combined results of Refs.~\cite{Franchi2017, Borkowski2017a} and we can use \emph{ab initio}-based $C_6(0_u^-)$ and $C_8(0_u^-)$ coefficients~\cite{Porsev2014} to calculate the bound state energies using a simple model potential $V(0_u^-) = -C_8(0_u^-)R^{-8}-(1-(\sigma/R)^6))C_6(0_u^-)R^{-6}$. We include the $-C_8(0_u^-)\,R^{-8}$ interaction for improved accuracy. $C_6(0_u^-)$ is taken directly from Ref.~\cite{Porsev2014} while $C_8(0_u^-)$ is estimated using $C_8(0_u^+)$ and $C_8(1_u)$, see Table~\ref{tbl:parameters} for details.  By setting $\sigma=9.0155455\,a_0$ we tune the scattering length of this potential to the combined experimental value of $a_{\rm ge}=94.84(0.14)\,a_0$~\cite{Franchi2017,Borkowski2017a} ($a_0 \approx 0.0529177~{\rm nm}$ is the Bohr radius).
	
	\begin{table}[b]
		\caption{Positions $E_e$ and error budget for three near-threshold vibrational states in the rotationless $^1$S$_0$+$^3$P$_0$ $0_u^-$ state of $^{174}$Yb$_2$ calculated using a model potential fitted to the $s$-wave scattering length $a_{\rm ge} = +94.84(0.14)a_0$ \cite{Franchi2017, Borkowski2017a}, the theoretical van der Waals coefficients $C_6=2561(95)E_h a_0^6$ \cite{Porsev2014} and the estimated $C_8 = 3.81(0.21)\times 10^5 E_h a_0^8$. See text for details.
			The total uncertainty is calculated from individual contributions from each of the parameters.  \label{tbl:positions}}
		\begin{ruledtabular}
			\begin{tabular}{l r r r}
				& $v_e'=-1$ & $v_e'=-2$ & $v_e'=-3$ \\
				\hline
				$E_e$ (MHz) & $-13.7554$ & $-322.4590$ & $-1438.7356$
				\\
				Uncertainty from $a_{\rm ge}$ & $0.0421 $ & $0.3356$ & $0.9103$ \\
				\hspace{10mm}''\hspace{10mm} $C_6$ & $0.0030$ & $4.1221$ & $22.4799$ \\
				\hspace{10mm}''\hspace{10mm} $C_8$ & $0.0066$ & $0.2471$ & $1.5109$ \\
				Total uncertainty & $0.0428$ & $4.1432$ & $22.5490$ \\
				Final & $-13.755(43)$ & $-322.5(4.2)$ & $-1439(23)$ \\
			\end{tabular}
		\end{ruledtabular}
	\end{table}	
	
We calculate the positions $E_e$ of excited state vibrational levels by 
numerically solving~\cite{Colbert1992, Tiesinga1998} the radial Schr\"odinger equation $\left(-(\hbar^2/2\mu)(d^2/dR^2) + V(0_u^-)\right)\Psi_e(R) = E_e \Psi_e(R)$. The reduced mass $\mu$ is half the mass of the $^{174}$Yb atom~\cite{Angeli2011}.
The energies of three most weakly bound $^1$S$_0$+$^3$P$_0$ states, and the appropriate error budgets, are given in Table~\ref{tbl:positions}. The energy of $-13.755(43)$~MHz for the $v'_e=-1$ state is determined to particularly high accuracy which will aid the initial search in a future experiment. The uncertainty of the theoretical $C_6(0_u^-)$ parameter has the largest impact on the positions of $v'_e=-2$ and $v'_e=-3$ states; the contribution from the $C_8$ parameter only plays a minor role in the total error budget of any of the reported states. 
The observation of transitions to the $v'_e=-1$ state alone would make it possible to constrain the $C_6(0_u^-)$ parameter and improve the predictions for more deeply bound states. 
The positions of the molecular clock lines  with respect to the atomic $^1$S$_0$$\rightarrow$$^3$P$_0$ line, shown in Fig.~\ref{fig:boundwf}a, depend on the difference between energies of excited and ground vibrational levels relative to their dissociation limits. The positions of three most weakly bound ground state vibrational levels have already been measured by two-color photoassociation spectroscopy: $-10.62513(53)$~MHz, $-325.66378(98)$~MHz, and $-1527.88543(34)$~MHz~\cite{Borkowski2017a}. 
The resulting molecular line positions are given in Table~\ref{tbl:fcf}. Of particular interest are the $-1\rightarrow-1$ and $-2\rightarrow-2$ transitions at $-3.130(43)$~MHz and $+3.2(4.2)$~MHz, which are only a few MHz away from the atomic clock line and which, as shown below, yield the best Rabi frequencies.
	
	In bosonic spin-singlet atoms the $^1$S$_0$$\rightarrow$$^3$P$_0$ transitions are strictly forbidden. Optical atomic clocks utilizing bosonic species rely on the technique pioneered in Refs.~\cite{Barber2006, Taichenachev2006} of applying small static magnetic fields $\mathbf B$ to induce (slight) Zeeman mixing of atomic ($m=0$) states 
	$\left|\,^3{\rm P}_0\right>$
	and 
	$\left|\,^3{\rm P}_1\right>$. The resulting mixed state 
	$\left|\,^3{\rm \widetilde P}_0\right> \approx \left|\,^3{\rm P}_0\right>+\frac{\Omega_B}{\Delta} \left|\,^3{\rm P}_1\right>$, where 
	$\Omega_B = \left<\,^3{\rm P}_0\right|\hat \mu \cdot \mathbf B \left|\,^3{\rm P}_1\right>/\hbar$ is the Zeeman matrix element, and $\Delta$ is the fine splitting between the two triplet states. The effective Rabi frequency of a clock transition interrogated by an optical field of amplitude $\mathbf E$ is 
	$\Omega_{\rm at} = \left<\,^1{\rm S}_0\right|\hat{\mathbf d} \cdot \mathbf E \left|\,^3{\rm \widetilde{P}}_0\right>/\hbar = \frac{\Omega_B}{\Delta}\left<\,^1{\rm S}_0\right|\hat{\mathbf d} \cdot \mathbf E \left|\,^3{\rm P}_1\right>/\hbar$. The atomic Rabi frequency, assuming parallel magnetic and optical fields, can be conveniently written as $\Omega_{\rm at} = \alpha \sqrt{I} |\mathbf B|$, where $I$ is the laser intensity. For Yb $\alpha=186\,{\rm Hz}/({\rm T\sqrt{\rm mW/cm^2}})$~\cite{Taichenachev2006}. We can also use this result for weakly bound molecules.
	It is typical for vibrational wavefunctions of near-threshold vibrational energy levels (Fig. \ref{fig:potentials}a) to extend to many tens of $a_0$~\cite{Jones2006}. For internuclear distances $R \gtrapprox 15\,a_0$ the atomic interaction in Yb$_2$ is small compared to the fine splitting $\Delta$ and may be treated as a perturbation. In the absence of external fields we can write the molecular state of indistinguishable atoms $A$ and $B$ as products of atomic states~\cite{Gao1996, Ciurylo2004, Mies1978}: the ground state $\left|g\right> = \left|\,^1{\rm S}_0\right>_A\left|\,^1{\rm S}_0\right>_B \Psi_g(R) Y_l^{m_l} (\theta, \phi)$ and the symmetrized \emph{ungerade} excited state 
	$\left|e\right> = \frac{1}{\sqrt{2}} \left( \left|\,^3{\rm P}_0\right>_A\left|\,^1{\rm S}_0\right>_B - (-1)^L \left|\,^1{\rm S}_0\right>_A\left|\,^3{\rm P}_0\right>_B \right) \Psi_e(R) Y_l^{m_l} (\theta, \phi)$. The total orbital angular momentum $L=1$; $\Psi_{g,e}(R)$ are the respective ground ($g$) and excited ($e$) radial wavefunctions, and $Y_l^{m_l}(\theta, \phi)$ are spherical harmonics describing the rotational motion. Here we only consider $s$-wave states with $l=m_l=0$. In presence of a magnetic field $\mathbf{B}$ the atomic states $\left|\,^3{\rm P}_0\right>_{A,B}$ in $\left|e\right>$ are replaced by $\left|\,^3{\rm \widetilde P}_0\right>_{A,B}$ forming the mixed molecular state~$\left|\tilde e \right>$. The molecular Rabi frequency is, by definition, $\Omega_{\rm mol} = \left<g\right|\hat{\mathbf d}_{\rm mol} \cdot \mathbf E \left|\,\tilde e\right>/\hbar$. For atoms at distances where the electron clouds are well separated, but still much closer than the transition wavelength $\lambda = 578\, {\rm nm}$, the molecular dipole moment $\hat{\mathbf d}_{\rm mol} = \hat{\mathbf d}_{A}+\hat{\mathbf d}_{B}$~\cite{Power1967}. Assuming the splitting $\Delta$ to be independent of $R$, we arrive at the intuitive result
	\begin{equation}
	\Omega_{\rm mol} = \sqrt{2} \sqrt{f_{\rm FC}}\, \Omega_{\rm at}, \label{eq:omega}
	\end{equation}
	where $f_{\rm FC} = \left|\int_0^{\infty} \Psi_g(R) \Psi_e(R) dR \right|^2$ is the usual Franck-Condon factor between ground- and excited-state wavefunctions $\Psi_g (R)$ and $\Psi_e (R)$. The $\sqrt{2}$ is a superradiant enhancement factor typical for $g$-$u$ transitions allowed by the Laporte rule. For Franck-Condon factors $f_{\rm FC}$ close to unity the molecular Rabi frequencies $\Omega_{\rm mol}$ could be comparable, or even slightly exceed, the atomic Rabi frequency $\Omega_{\rm at}$. It should be noted that for short distances $R$ this simple derivation no longer holds and magnetic couplings between the $^1$S$_0$+$^3$P$_0$ $0_u^-$, and the two $^1$S$_0$+$^3$P$_1$ $0_u^+$, $1_u$ states should be considered instead.

We have investigated the possibility of direct clock photoassociation of unbound atomic pairs in  3D optical lattice sites. 
The result is negative: even for a strong lattice with lattice frequency $\omega = 100\,{\rm kHz}$ the respective Franck-Condon factors for $v'_e=-1,-2,-3$ are $1.06\times 10^{-5}$, $1.49\times 10^{-5}$ and $1.14\times 10^{-5}$. These, combined with the low atomic transition probability would make direct photoassociation experimentally challenging. In principle, stronger magnetic and optical fields could be used to compensate for the small $f_{\rm FC}$. For example, in a lattice with an experimentally reasonable $\omega = 20~{\rm kHz}$ a Rabi frequency of only 1~Hz for a transition to the $v_e=-1$ state would require a laser intensity  of 150~W/cm$^2$ even with the magnetic field increased to $10\,{\rm mT}$. This may explain why no experimental observation of clock line photoassociation has been so far reported.

\begin{table}[b]
	\caption{Positions $\Delta \nu$ of molecular clock lines relative to the atomic clock transition. The respective Franck-Condon factors $f_{FC}$ were calculated using the ground state potential of Ref.~\cite{Borkowski2017a} and the present $0_u^-$ interaction model. The uncertainties are dominated by the error bars of the excited state energy level positions (see Table~\ref{tbl:positions}). \label{tbl:fcf}}
	\begin{ruledtabular}
		\begin{tabular}{l r r r r r r }
			& \multicolumn{2}{c}{$v'_e=-1$} & \multicolumn{2}{c}{$v'_e=-2$}& \multicolumn{2}{c}{$v'_e=-3$}\\
			\cline{2-3} \cline{4-5} \cline{6-7}
			$v'_g$ & $\Delta \nu$ (MHz) & $f_{\rm FC}$ & $\Delta \nu$ (MHz) & $f_{\rm FC}$ & $\Delta \nu$ (MHz) & $f_{\rm FC}$ \\
			\cline{1-2}
			\hline
			$-1$ & $-3.130(43)$ & $0.9495$ & $-311.8(4.2)$ & $0.0217$ & $-1428(23)$ & $0.0166$ \\
			$-2$ & $311.908(43)$& $0.0009$ & $3.2(4.2)$ & $0.6755$ & $-1113(23)$ & $0.2140$ \\
			$-3$ & $1514.130(43)$& $0.0002$ & $1205.4(4.2)$ & $0.0025$ & $89(23)$ & $0.3990$ \\
		\end{tabular}
	\end{ruledtabular}
\end{table}		

The Franck-Condon factors can be improved by orders of magnitude if transitions between weakly bound $^1$S$_0$+$^1$S$_0$ and $^1$S$_0$+$^3$P$_0$ states are used instead of direct photoassociation.
The positions of near-threshold energy levels with the same vibrational quantum numbers $v'_{g,e}$ are very similar and, consequently, molecular transitions lie close to the atomic line (Fig.~\ref{fig:boundwf}a). The respective $C_6$ factors being comparable ($C_6(0_g^+)\approx 1937 \, E_h a_0^6$ \cite{Borkowski2017a}), the classical outer turning points are also similar and the radial wavefunctions (Fig.~\ref{fig:boundwf}b) have significant overlap. This leads to unusually favorable Franck-Condon factors reaching even $0.95$ for the $-1\rightarrow-1$ transition, see Table~\ref{tbl:fcf}. In Fig.~\ref{fig:boundwf}c we show projected molecular Rabi frequencies as a function of laser intensity $I$ for a magnetic field $|\mathbf B|=1\,{\rm mT}$. As expected, they are comparable to $\Omega_{\rm at}$: values of a few Hz can be obtained for laser intensities well below $1\,{\rm W/cm^2}$. In a practical experimental realization, the same laser power (and possibly even the same beam) could be used for atomic and molecular transitions. 
	
Weakly bound ground state $^{174}$Yb$_2$ molecules necessary for clock spectroscopy could be produced by STIRAP~\cite{Bergmann1998,Vitanov1997,Winkler2007,Stellmer2012, Ciamei2017} in a deep optical lattice using $^1$S$_0$+$^3$P$_1$ $0_u^+$ energy levels as intermediate states. The Franck-Condon factors for the STIRAP ``pump'' and ``dump'' transitions are not nearly as convenient as in, for example, strontium due to extended outer turning points in the excited state. We have, however, tentatively checked by numerical simulations of the three level dynamics~\cite{Ciamei2017} that optical production of weakly bound Yb molecules is still possible. 
Ground state molecules in the $v_g=-1$ state could be produced using the known $0_u^+$ $-1974$-MHz energy level~\cite{Borkowski2009} as an intermediate state.
STIRAP pulses executed over 100~ms in a lattice of trapping frequency $\omega=20\,{\rm kHz}$, and with moderate maximum pump and dump laser intensities of $\rm 50\,W/cm^2$ and $\rm 10\,mW/cm^2$ would yield about 75\% transfer efficiency. For the $v_g=-2$ state a 69\% efficiency is obtained under the same conditions apart from the dump laser intensity being increased to 1~$\rm W/cm^2$. For the $v_g=-3$ state a deeper intermediate state at $-3678$~MHz would have to be used and for both maximum laser intensities equal to 50~$\rm W/cm^2$ would give an efficiency of 51\%. Given the high efficiencies, STIRAP could also be used to dissociate~\cite{Ciamei2017} the remaining ground state molecules for detection after clock line interrogation. In a thermal gas, molecules in the most weakly bound $v_g=-1$ state could also be produced by spontaneous decay in photoassociation~\cite{Fioretti1998, Reinaudi2012} on the $-1974$-MHz line at an efficiency of about~6\%. 
	
So far we have ignored the role of short range molecular physics. In principle, strong coupling with radiative molecular states could shorten the lifetimes of the $0_u^-$ states. Out of molecular states correlating with $^1$S$_0$+$^3$P$_{1,2}$ and $^1$S$_0$+$^1$P$_1$ asymptotes the only spin-orbit coupling occurs with the higher lying metastable $^1$S$_0$+$^3$P$_2$ $0_u^-$ state~\cite{Mies1978}. This way the $^1$S$_0$+$^3$P$_0$ state is the lowest state of $0_u^-$ symmetry and as such would be immune to predissociation, which limited the lifetimes of e.g. the subradiant states in Sr$_2$~\cite{McGuyer2015}. It is possible that complications could arise from other, high lying, atomic asymptotes, e.g. $^1$S$_0$+$^3$D$_{1,2,3}$; unfortunately, no \emph{ab initio} calculations of these states are available at this point and it is not known if any of related molecular potentials would cross the $^1$S$_0$+$^3$P$_0$ dissociation limit. We note, however, that both Bouganne~\emph{et al.}~\cite{Bouganne2017} and Franchi~\emph{et al.}~\cite{Franchi2017} report a lack of inelastic $g$-$e$ collisions, hinting at long lived molecular states. Another possible consequence of short range molecular interactions is the possible enhancement of $\Omega_{\rm mol}$ from short range variation of the splitting $\Delta$ or if a perturber $0_u^+$ or $1_u$ vibrational state accidentally coincides with the probed energy level. If present, such effects could be compensated for by dialing down the magnetic field.
	
We have shown that spectroscopy of weakly bound cold molecules near the $^1$S$_0$$\rightarrow$$^3$P$_0$ clock transition may be experimentally feasible. We have predicted, to high accuracy, the positions of three most weakly bound $0_u^-$ energy levels below the $^1$S$_0$+$^3$P$_0$ asymptote of the bosonic $^{174}$Yb$_2$ molecule. We have shown that thanks to favorable Franck-Condon factors a molecular analog of magnetic-field-induced optical clock spectroscopy~\cite{Taichenachev2006, Barber2006} is possible. The magnetically induced Rabi frequencies for optical clock transitions between weakly bound $^1$S$_0$+$^1$S$_0$ $0_g^+$ and $^1$S$_0$+$^3$P$_0$ $0_u^-$ energy levels can be comparable to those for the atomic clock transition. The necessary ground-state weakly bound molecules could be produced by either spontaneous emission following intercombination line photoassociation, or by the STIRAP technique in deep optical lattice utilizing one of the known $^1$S$_0$+$^3$P$_1$ $0_u^+$ intermediate states. The predicted transitions could be observed in existing optical atomic clock setups that use $^{174}$Yb atoms. Since the $^1$S$_0$+$^3$P$_0$ $0_u^-$ state is immune to predissociation, its vibrational states could be long lived and their positions could be measured at the hertz level. The two strongest transitions lie at very small detunings with respect to the atomic clock line: $-3.130(43)$~MHz and $+3.2(4.2)$~MHz. In principle, depending on the prepared sample (whether atomic or molecular, and in which vibrational state), the same experimental apparatus could be used to compare the readings of an atomic and molecular frequency standard. Possible applications include searching for the temporal variation of the fundamental constants~\cite{Lea2007, Blatt2008, Shelkovnikov2008, Beloy2011, Orzel2012} or new physics beyond the Standard Model~\cite{Ferreira2012, Derevianko2014}. The mechanism is universal and should apply to other bosonic species used in optical clocks, like $^{88}$Sr. 
	
	\begin{acknowledgments}
	I would like to thank Roman Ciuryło and Piotr Żuchowski for carefully reading the manuscript and Marianna Safronova for useful discussions. This research was supported by the National Science Centre, Grant no. 2014/13/N/ST2/02591. 
	Support has been received from the project EMPIR 15SIB03 OC18. This project has received funding from the EMPIR programme co-financed by the Participating States and from the European Union’s Horizon 2020 research and innovation programme. It is part of an ongoing research program of the National Laboratory FAMO in Toru\'n, Poland. Calculations have been carried out in Wroclaw Centre for Networking and Supercomputing, Grant no. 353.
	\end{acknowledgments}
	\bibliography{library}

\begin{thebibliography}{62}%
\makeatletter
\providecommand \@ifxundefined [1]{%
 \@ifx{#1\undefined}
}%
\providecommand \@ifnum [1]{%
 \ifnum #1\expandafter \@firstoftwo
 \else \expandafter \@secondoftwo
 \fi
}%
\providecommand \@ifx [1]{%
 \ifx #1\expandafter \@firstoftwo
 \else \expandafter \@secondoftwo
 \fi
}%
\providecommand \natexlab [1]{#1}%
\providecommand \enquote  [1]{``#1''}%
\providecommand \bibnamefont  [1]{#1}%
\providecommand \bibfnamefont [1]{#1}%
\providecommand \citenamefont [1]{#1}%
\providecommand \href@noop [0]{\@secondoftwo}%
\providecommand \href [0]{\begingroup \@sanitize@url \@href}%
\providecommand \@href[1]{\@@startlink{#1}\@@href}%
\providecommand \@@href[1]{\endgroup#1\@@endlink}%
\providecommand \@sanitize@url [0]{\catcode `\\12\catcode `\$12\catcode
  `\&12\catcode `\#12\catcode `\^12\catcode `\_12\catcode `\%12\relax}%
\providecommand \@@startlink[1]{}%
\providecommand \@@endlink[0]{}%
\providecommand \url  [0]{\begingroup\@sanitize@url \@url }%
\providecommand \@url [1]{\endgroup\@href {#1}{\urlprefix }}%
\providecommand \urlprefix  [0]{URL }%
\providecommand \Eprint [0]{\href }%
\providecommand \doibase [0]{http://dx.doi.org/}%
\providecommand \selectlanguage [0]{\@gobble}%
\providecommand \bibinfo  [0]{\@secondoftwo}%
\providecommand \bibfield  [0]{\@secondoftwo}%
\providecommand \translation [1]{[#1]}%
\providecommand \BibitemOpen [0]{}%
\providecommand \bibitemStop [0]{}%
\providecommand \bibitemNoStop [0]{.\EOS\space}%
\providecommand \EOS [0]{\spacefactor3000\relax}%
\providecommand \BibitemShut  [1]{\csname bibitem#1\endcsname}%
\let\auto@bib@innerbib\@empty
\bibitem [{\citenamefont {Katori}(2011)}]{Katori2011}%
  \BibitemOpen
  \bibfield  {author} {\bibinfo {author} {\bibfnamefont {H.}~\bibnamefont
  {Katori}},\ }\href {\doibase 10.1038/nphoton.2011.45} {\bibfield  {journal}
  {\bibinfo  {journal} {Nat. Photonics}\ }\textbf {\bibinfo {volume} {5}},\
  \bibinfo {pages} {203} (\bibinfo {year} {2011})}\BibitemShut {NoStop}%
\bibitem [{\citenamefont {Ludlow}\ \emph {et~al.}(2015)\citenamefont {Ludlow},
  \citenamefont {Boyd}, \citenamefont {Ye}, \citenamefont {Peik},\ and\
  \citenamefont {Schmidt}}]{Ludlow2015}%
  \BibitemOpen
  \bibfield  {author} {\bibinfo {author} {\bibfnamefont {A.~D.}\ \bibnamefont
  {Ludlow}}, \bibinfo {author} {\bibfnamefont {M.~M.}\ \bibnamefont {Boyd}},
  \bibinfo {author} {\bibfnamefont {J.}~\bibnamefont {Ye}}, \bibinfo {author}
  {\bibfnamefont {E.}~\bibnamefont {Peik}}, \ and\ \bibinfo {author}
  {\bibfnamefont {P.~O.}\ \bibnamefont {Schmidt}},\ }\href {\doibase
  10.1103/RevModPhys.87.637} {\bibfield  {journal} {\bibinfo  {journal} {Rev.
  Mod. Phys.}\ }\textbf {\bibinfo {volume} {87}},\ \bibinfo {pages} {638}
  (\bibinfo {year} {2015})}\BibitemShut {NoStop}%
\bibitem [{\citenamefont {Chou}\ \emph {et~al.}(2010)\citenamefont {Chou},
  \citenamefont {Hume}, \citenamefont {Rosenband},\ and\ \citenamefont
  {Wineland}}]{Chou2010}%
  \BibitemOpen
  \bibfield  {author} {\bibinfo {author} {\bibfnamefont {C.~W.}\ \bibnamefont
  {Chou}}, \bibinfo {author} {\bibfnamefont {D.~B.}\ \bibnamefont {Hume}},
  \bibinfo {author} {\bibfnamefont {T.}~\bibnamefont {Rosenband}}, \ and\
  \bibinfo {author} {\bibfnamefont {D.~J.}\ \bibnamefont {Wineland}},\ }\href
  {\doibase 10.1126/science.1192720} {\bibfield  {journal} {\bibinfo  {journal}
  {Science (80-. ).}\ }\textbf {\bibinfo {volume} {329}},\ \bibinfo {pages}
  {1630} (\bibinfo {year} {2010})}\BibitemShut {NoStop}%
\bibitem [{\citenamefont {Lea}(2007)}]{Lea2007}%
  \BibitemOpen
  \bibfield  {author} {\bibinfo {author} {\bibfnamefont {S.~N.}\ \bibnamefont
  {Lea}},\ }\href {\doibase 10.1088/0034-4885/70/9/R01} {\bibfield  {journal}
  {\bibinfo  {journal} {Rep. Prog. Phys.}\ }\textbf {\bibinfo {volume} {70}},\
  \bibinfo {pages} {1473} (\bibinfo {year} {2007})}\BibitemShut {NoStop}%
\bibitem [{\citenamefont {Fortier}\ \emph {et~al.}(2007)\citenamefont
  {Fortier}, \citenamefont {Ashby}, \citenamefont {Bergquist}, \citenamefont
  {Delaney}, \citenamefont {Diddams}, \citenamefont {Heavner}, \citenamefont
  {Hollberg}, \citenamefont {Itano}, \citenamefont {Jefferts}, \citenamefont
  {Kim}, \citenamefont {Levi}, \citenamefont {Lorini}, \citenamefont {Oskay},
  \citenamefont {Parker}, \citenamefont {Shirley},\ and\ \citenamefont
  {Stalnaker}}]{Fortier2007}%
  \BibitemOpen
  \bibfield  {author} {\bibinfo {author} {\bibfnamefont {T.~M.}\ \bibnamefont
  {Fortier}}, \bibinfo {author} {\bibfnamefont {N.}~\bibnamefont {Ashby}},
  \bibinfo {author} {\bibfnamefont {J.~C.}\ \bibnamefont {Bergquist}}, \bibinfo
  {author} {\bibfnamefont {M.~J.}\ \bibnamefont {Delaney}}, \bibinfo {author}
  {\bibfnamefont {S.~A.}\ \bibnamefont {Diddams}}, \bibinfo {author}
  {\bibfnamefont {T.~P.}\ \bibnamefont {Heavner}}, \bibinfo {author}
  {\bibfnamefont {L.}~\bibnamefont {Hollberg}}, \bibinfo {author}
  {\bibfnamefont {W.~M.}\ \bibnamefont {Itano}}, \bibinfo {author}
  {\bibfnamefont {S.~R.}\ \bibnamefont {Jefferts}}, \bibinfo {author}
  {\bibfnamefont {K.}~\bibnamefont {Kim}}, \bibinfo {author} {\bibfnamefont
  {F.}~\bibnamefont {Levi}}, \bibinfo {author} {\bibfnamefont {L.}~\bibnamefont
  {Lorini}}, \bibinfo {author} {\bibfnamefont {W.~H.}\ \bibnamefont {Oskay}},
  \bibinfo {author} {\bibfnamefont {T.~E.}\ \bibnamefont {Parker}}, \bibinfo
  {author} {\bibfnamefont {J.}~\bibnamefont {Shirley}}, \ and\ \bibinfo
  {author} {\bibfnamefont {J.~E.}\ \bibnamefont {Stalnaker}},\ }\href {\doibase
  10.1103/PhysRevLett.98.070801} {\bibfield  {journal} {\bibinfo  {journal}
  {Phys. Rev. Lett.}\ }\textbf {\bibinfo {volume} {98}},\ \bibinfo {pages}
  {070801} (\bibinfo {year} {2007})}\BibitemShut {NoStop}%
\bibitem [{\citenamefont {Blatt}\ \emph {et~al.}(2008)\citenamefont {Blatt},
  \citenamefont {Ludlow}, \citenamefont {Campbell}, \citenamefont {Thomsen},
  \citenamefont {Zelevinsky}, \citenamefont {Boyd}, \citenamefont {Ye},
  \citenamefont {Baillard}, \citenamefont {Fouch{\'{e}}}, \citenamefont {{Le
  Targat}}, \citenamefont {Brusch}, \citenamefont {Lemonde}, \citenamefont
  {Takamoto}, \citenamefont {Hong}, \citenamefont {Katori},\ and\ \citenamefont
  {Flambaum}}]{Blatt2008}%
  \BibitemOpen
  \bibfield  {author} {\bibinfo {author} {\bibfnamefont {S.}~\bibnamefont
  {Blatt}}, \bibinfo {author} {\bibfnamefont {A.~D.}\ \bibnamefont {Ludlow}},
  \bibinfo {author} {\bibfnamefont {G.~K.}\ \bibnamefont {Campbell}}, \bibinfo
  {author} {\bibfnamefont {J.~W.}\ \bibnamefont {Thomsen}}, \bibinfo {author}
  {\bibfnamefont {T.}~\bibnamefont {Zelevinsky}}, \bibinfo {author}
  {\bibfnamefont {M.~M.}\ \bibnamefont {Boyd}}, \bibinfo {author}
  {\bibfnamefont {J.}~\bibnamefont {Ye}}, \bibinfo {author} {\bibfnamefont
  {X.}~\bibnamefont {Baillard}}, \bibinfo {author} {\bibfnamefont
  {M.}~\bibnamefont {Fouch{\'{e}}}}, \bibinfo {author} {\bibfnamefont
  {R.}~\bibnamefont {{Le Targat}}}, \bibinfo {author} {\bibfnamefont
  {A.}~\bibnamefont {Brusch}}, \bibinfo {author} {\bibfnamefont
  {P.}~\bibnamefont {Lemonde}}, \bibinfo {author} {\bibfnamefont
  {M.}~\bibnamefont {Takamoto}}, \bibinfo {author} {\bibfnamefont {F.~L.}\
  \bibnamefont {Hong}}, \bibinfo {author} {\bibfnamefont {H.}~\bibnamefont
  {Katori}}, \ and\ \bibinfo {author} {\bibfnamefont {V.~V.}\ \bibnamefont
  {Flambaum}},\ }\href {\doibase 10.1103/PhysRevLett.100.140801} {\bibfield
  {journal} {\bibinfo  {journal} {Phys. Rev. Lett.}\ }\textbf {\bibinfo
  {volume} {100}},\ \bibinfo {pages} {140801} (\bibinfo {year}
  {2008})}\BibitemShut {NoStop}%
\bibitem [{\citenamefont {Orzel}(2012)}]{Orzel2012}%
  \BibitemOpen
  \bibfield  {author} {\bibinfo {author} {\bibfnamefont {C.}~\bibnamefont
  {Orzel}},\ }\href {\doibase 10.1088/0031-8949/86/06/068101} {\bibfield
  {journal} {\bibinfo  {journal} {Phys. Scr.}\ }\textbf {\bibinfo {volume}
  {86}},\ \bibinfo {pages} {68101} (\bibinfo {year} {2012})}\BibitemShut
  {NoStop}%
\bibitem [{\citenamefont {Ferreira}\ \emph {et~al.}(2012)\citenamefont
  {Ferreira}, \citenamefont {Juli{\~{a}}o}, \citenamefont {Martins},\ and\
  \citenamefont {Monteiro}}]{Ferreira2012}%
  \BibitemOpen
  \bibfield  {author} {\bibinfo {author} {\bibfnamefont {M.~C.}\ \bibnamefont
  {Ferreira}}, \bibinfo {author} {\bibfnamefont {M.~D.}\ \bibnamefont
  {Juli{\~{a}}o}}, \bibinfo {author} {\bibfnamefont {C.~J.}\ \bibnamefont
  {Martins}}, \ and\ \bibinfo {author} {\bibfnamefont {A.~M.}\ \bibnamefont
  {Monteiro}},\ }\href {\doibase 10.1103/PhysRevD.86.125025} {\bibfield
  {journal} {\bibinfo  {journal} {Phys. Rev. D}\ }\textbf {\bibinfo {volume}
  {86}},\ \bibinfo {pages} {125025} (\bibinfo {year} {2012})}\BibitemShut
  {NoStop}%
\bibitem [{\citenamefont {Safronova}\ \emph {et~al.}(2017)\citenamefont
  {Safronova}, \citenamefont {Budker}, \citenamefont {DeMille}, \citenamefont
  {Kimball}, \citenamefont {Derevianko},\ and\ \citenamefont
  {Clark}}]{Safronova2017}%
  \BibitemOpen
  \bibfield  {author} {\bibinfo {author} {\bibfnamefont {M.~S.}\ \bibnamefont
  {Safronova}}, \bibinfo {author} {\bibfnamefont {D.}~\bibnamefont {Budker}},
  \bibinfo {author} {\bibfnamefont {D.}~\bibnamefont {DeMille}}, \bibinfo
  {author} {\bibfnamefont {D.~F.~J.}\ \bibnamefont {Kimball}}, \bibinfo
  {author} {\bibfnamefont {A.}~\bibnamefont {Derevianko}}, \ and\ \bibinfo
  {author} {\bibfnamefont {C.~W.}\ \bibnamefont {Clark}},\ }\href
  {http://arxiv.org/abs/1710.01833} {\  (\bibinfo {year} {2017})},\ \Eprint
  {http://arxiv.org/abs/1710.01833} {arXiv:1710.01833} \BibitemShut {NoStop}%
\bibitem [{\citenamefont {Martin}\ \emph {et~al.}(2013)\citenamefont {Martin},
  \citenamefont {Bishof}, \citenamefont {Swallows}, \citenamefont {Zhang},
  \citenamefont {Benko}, \citenamefont {Von-Stecher}, \citenamefont
  {Gorschkow}, \citenamefont {Rey},\ and\ \citenamefont {Ye}}]{Martin2013}%
  \BibitemOpen
  \bibfield  {author} {\bibinfo {author} {\bibfnamefont {M.~J.}\ \bibnamefont
  {Martin}}, \bibinfo {author} {\bibfnamefont {M.}~\bibnamefont {Bishof}},
  \bibinfo {author} {\bibfnamefont {M.~D.}\ \bibnamefont {Swallows}}, \bibinfo
  {author} {\bibfnamefont {X.}~\bibnamefont {Zhang}}, \bibinfo {author}
  {\bibfnamefont {C.}~\bibnamefont {Benko}}, \bibinfo {author} {\bibfnamefont
  {J.}~\bibnamefont {Von-Stecher}}, \bibinfo {author} {\bibfnamefont {A.~V.}\
  \bibnamefont {Gorschkow}}, \bibinfo {author} {\bibfnamefont {A.~M.}\
  \bibnamefont {Rey}}, \ and\ \bibinfo {author} {\bibfnamefont
  {J.}~\bibnamefont {Ye}},\ }\href {\doibase 10.1126/science.1236929}
  {\bibfield  {journal} {\bibinfo  {journal} {Science (80-. ).}\ }\textbf
  {\bibinfo {volume} {341}},\ \bibinfo {pages} {632} (\bibinfo {year}
  {2013})}\BibitemShut {NoStop}%
\bibitem [{\citenamefont {Rey}\ \emph {et~al.}(2014)\citenamefont {Rey},
  \citenamefont {Gorshkov}, \citenamefont {Kraus}, \citenamefont {Martin},
  \citenamefont {Bishof}, \citenamefont {Zhang}, \citenamefont {Benko},
  \citenamefont {Ye}, \citenamefont {Lemke},\ and\ \citenamefont
  {Ludlow}}]{Rey2014}%
  \BibitemOpen
  \bibfield  {author} {\bibinfo {author} {\bibfnamefont {A.}~\bibnamefont
  {Rey}}, \bibinfo {author} {\bibfnamefont {A.}~\bibnamefont {Gorshkov}},
  \bibinfo {author} {\bibfnamefont {C.}~\bibnamefont {Kraus}}, \bibinfo
  {author} {\bibfnamefont {M.}~\bibnamefont {Martin}}, \bibinfo {author}
  {\bibfnamefont {M.}~\bibnamefont {Bishof}}, \bibinfo {author} {\bibfnamefont
  {X.}~\bibnamefont {Zhang}}, \bibinfo {author} {\bibfnamefont
  {C.}~\bibnamefont {Benko}}, \bibinfo {author} {\bibfnamefont
  {J.}~\bibnamefont {Ye}}, \bibinfo {author} {\bibfnamefont {N.}~\bibnamefont
  {Lemke}}, \ and\ \bibinfo {author} {\bibfnamefont {A.}~\bibnamefont
  {Ludlow}},\ }\href {\doibase 10.1016/j.aop.2013.11.002} {\bibfield  {journal}
  {\bibinfo  {journal} {Ann. Phys. (N. Y).}\ }\textbf {\bibinfo {volume}
  {340}},\ \bibinfo {pages} {311} (\bibinfo {year} {2014})}\BibitemShut
  {NoStop}%
\bibitem [{\citenamefont {Dareau}\ \emph {et~al.}(2015)\citenamefont {Dareau},
  \citenamefont {Scholl}, \citenamefont {Beaufils}, \citenamefont
  {D{\"{o}}ring}, \citenamefont {Beugnon},\ and\ \citenamefont
  {Gerbier}}]{Dareau2015}%
  \BibitemOpen
  \bibfield  {author} {\bibinfo {author} {\bibfnamefont {A.}~\bibnamefont
  {Dareau}}, \bibinfo {author} {\bibfnamefont {M.}~\bibnamefont {Scholl}},
  \bibinfo {author} {\bibfnamefont {Q.}~\bibnamefont {Beaufils}}, \bibinfo
  {author} {\bibfnamefont {D.}~\bibnamefont {D{\"{o}}ring}}, \bibinfo {author}
  {\bibfnamefont {J.}~\bibnamefont {Beugnon}}, \ and\ \bibinfo {author}
  {\bibfnamefont {F.}~\bibnamefont {Gerbier}},\ }\href {\doibase
  10.1103/PhysRevA.91.023626} {\bibfield  {journal} {\bibinfo  {journal} {Phys.
  Rev. A}\ }\textbf {\bibinfo {volume} {91}},\ \bibinfo {pages} {023626}
  (\bibinfo {year} {2015})}\BibitemShut {NoStop}%
\bibitem [{\citenamefont {Derevianko}\ and\ \citenamefont
  {Pospelov}(2014)}]{Derevianko2014}%
  \BibitemOpen
  \bibfield  {author} {\bibinfo {author} {\bibfnamefont {A.}~\bibnamefont
  {Derevianko}}\ and\ \bibinfo {author} {\bibfnamefont {M.}~\bibnamefont
  {Pospelov}},\ }\href {\doibase 10.1038/nphys3137} {\bibfield  {journal}
  {\bibinfo  {journal} {Nat. Phys.}\ }\textbf {\bibinfo {volume} {10}},\
  \bibinfo {pages} {933} (\bibinfo {year} {2014})}\BibitemShut {NoStop}%
\bibitem [{\citenamefont {Wcis{\l}o}\ \emph {et~al.}(2016)\citenamefont
  {Wcis{\l}o}, \citenamefont {Morzy{\'{n}}ski}, \citenamefont {Bober},
  \citenamefont {Cygan}, \citenamefont {Lisak}, \citenamefont {Ciury{\l}o},\
  and\ \citenamefont {Zawada}}]{Wcislo2016}%
  \BibitemOpen
  \bibfield  {author} {\bibinfo {author} {\bibfnamefont {P.}~\bibnamefont
  {Wcis{\l}o}}, \bibinfo {author} {\bibfnamefont {P.}~\bibnamefont
  {Morzy{\'{n}}ski}}, \bibinfo {author} {\bibfnamefont {M.}~\bibnamefont
  {Bober}}, \bibinfo {author} {\bibfnamefont {A.}~\bibnamefont {Cygan}},
  \bibinfo {author} {\bibfnamefont {D.}~\bibnamefont {Lisak}}, \bibinfo
  {author} {\bibfnamefont {R.}~\bibnamefont {Ciury{\l}o}}, \ and\ \bibinfo
  {author} {\bibfnamefont {M.}~\bibnamefont {Zawada}},\ }\href {\doibase
  10.1038/s41550-016-0009} {\bibfield  {journal} {\bibinfo  {journal} {Nat.
  Astron.}\ }\textbf {\bibinfo {volume} {1}},\ \bibinfo {pages} {0009}
  (\bibinfo {year} {2016})}\BibitemShut {NoStop}%
\bibitem [{\citenamefont {Shelkovnikov}\ \emph {et~al.}(2008)\citenamefont
  {Shelkovnikov}, \citenamefont {Butcher}, \citenamefont {Chardonnet},\ and\
  \citenamefont {Amy-Klein}}]{Shelkovnikov2008}%
  \BibitemOpen
  \bibfield  {author} {\bibinfo {author} {\bibfnamefont {A.}~\bibnamefont
  {Shelkovnikov}}, \bibinfo {author} {\bibfnamefont {R.~J.}\ \bibnamefont
  {Butcher}}, \bibinfo {author} {\bibfnamefont {C.}~\bibnamefont {Chardonnet}},
  \ and\ \bibinfo {author} {\bibfnamefont {A.}~\bibnamefont {Amy-Klein}},\
  }\href {\doibase 10.1103/PhysRevLett.100.150801} {\bibfield  {journal}
  {\bibinfo  {journal} {Phys. Rev. Lett.}\ }\textbf {\bibinfo {volume} {100}},\
  \bibinfo {pages} {150801} (\bibinfo {year} {2008})}\BibitemShut {NoStop}%
\bibitem [{\citenamefont {Zelevinsky}\ \emph {et~al.}(2008)\citenamefont
  {Zelevinsky}, \citenamefont {Kotochigova},\ and\ \citenamefont
  {Ye}}]{Zelevinsky2008}%
  \BibitemOpen
  \bibfield  {author} {\bibinfo {author} {\bibfnamefont {T.}~\bibnamefont
  {Zelevinsky}}, \bibinfo {author} {\bibfnamefont {S.}~\bibnamefont
  {Kotochigova}}, \ and\ \bibinfo {author} {\bibfnamefont {J.}~\bibnamefont
  {Ye}},\ }\href {\doibase 10.1103/PhysRevLett.100.043201} {\bibfield
  {journal} {\bibinfo  {journal} {Phys. Rev. Lett.}\ }\textbf {\bibinfo
  {volume} {100}},\ \bibinfo {pages} {043201} (\bibinfo {year}
  {2008})}\BibitemShut {NoStop}%
\bibitem [{\citenamefont {Demille}\ \emph {et~al.}(2008)\citenamefont
  {Demille}, \citenamefont {Sainis}, \citenamefont {Sage}, \citenamefont
  {Bergeman}, \citenamefont {Kotochigova},\ and\ \citenamefont
  {Tiesinga}}]{DeMille2008}%
  \BibitemOpen
  \bibfield  {author} {\bibinfo {author} {\bibfnamefont {D.}~\bibnamefont
  {Demille}}, \bibinfo {author} {\bibfnamefont {S.}~\bibnamefont {Sainis}},
  \bibinfo {author} {\bibfnamefont {J.}~\bibnamefont {Sage}}, \bibinfo {author}
  {\bibfnamefont {T.}~\bibnamefont {Bergeman}}, \bibinfo {author}
  {\bibfnamefont {S.}~\bibnamefont {Kotochigova}}, \ and\ \bibinfo {author}
  {\bibfnamefont {E.}~\bibnamefont {Tiesinga}},\ }\href {\doibase
  10.1103/PhysRevLett.100.043202} {\bibfield  {journal} {\bibinfo  {journal}
  {Phys. Rev. Lett.}\ }\textbf {\bibinfo {volume} {100}},\ \bibinfo {pages}
  {043202} (\bibinfo {year} {2008})}\BibitemShut {NoStop}%
\bibitem [{\citenamefont {Beloy}\ \emph {et~al.}(2011)\citenamefont {Beloy},
  \citenamefont {Hauser}, \citenamefont {Borschevsky}, \citenamefont
  {Flambaum},\ and\ \citenamefont {Schwerdtfeger}}]{Beloy2011}%
  \BibitemOpen
  \bibfield  {author} {\bibinfo {author} {\bibfnamefont {K.}~\bibnamefont
  {Beloy}}, \bibinfo {author} {\bibfnamefont {A.~W.}\ \bibnamefont {Hauser}},
  \bibinfo {author} {\bibfnamefont {A.}~\bibnamefont {Borschevsky}}, \bibinfo
  {author} {\bibfnamefont {V.~V.}\ \bibnamefont {Flambaum}}, \ and\ \bibinfo
  {author} {\bibfnamefont {P.}~\bibnamefont {Schwerdtfeger}},\ }\href {\doibase
  10.1103/PhysRevA.84.062114} {\bibfield  {journal} {\bibinfo  {journal} {Phys.
  Rev. A}\ }\textbf {\bibinfo {volume} {84}},\ \bibinfo {pages} {062114}
  (\bibinfo {year} {2011})}\BibitemShut {NoStop}%
\bibitem [{\citenamefont {Stadnik}\ and\ \citenamefont
  {Flambaum}(2015)}]{Stadnik2015}%
  \BibitemOpen
  \bibfield  {author} {\bibinfo {author} {\bibfnamefont {Y.~V.}\ \bibnamefont
  {Stadnik}}\ and\ \bibinfo {author} {\bibfnamefont {V.~V.}\ \bibnamefont
  {Flambaum}},\ }\href {\doibase 10.1103/PhysRevLett.115.201301} {\bibfield
  {journal} {\bibinfo  {journal} {Phys. Rev. Lett.}\ }\textbf {\bibinfo
  {volume} {115}},\ \bibinfo {pages} {201301} (\bibinfo {year} {2015})},\
  \Eprint {http://arxiv.org/abs/1503.08540} {arXiv:1503.08540} \BibitemShut
  {NoStop}%
\bibitem [{\citenamefont {Arvanitaki}\ \emph {et~al.}(2015)\citenamefont
  {Arvanitaki}, \citenamefont {Huang},\ and\ \citenamefont {{Van
  Tilburg}}}]{Arvanitaki2015}%
  \BibitemOpen
  \bibfield  {author} {\bibinfo {author} {\bibfnamefont {A.}~\bibnamefont
  {Arvanitaki}}, \bibinfo {author} {\bibfnamefont {J.}~\bibnamefont {Huang}}, \
  and\ \bibinfo {author} {\bibfnamefont {K.}~\bibnamefont {{Van Tilburg}}},\
  }\href {\doibase 10.1103/PhysRevD.91.015015} {\bibfield  {journal} {\bibinfo
  {journal} {Phys. Rev. D}\ }\textbf {\bibinfo {volume} {91}},\ \bibinfo
  {pages} {015015} (\bibinfo {year} {2015})}\BibitemShut {NoStop}%
\bibitem [{\citenamefont {Zhang}\ and\ \citenamefont
  {Dalgarno}(2008)}]{Zhang2008}%
  \BibitemOpen
  \bibfield  {author} {\bibinfo {author} {\bibfnamefont {P.}~\bibnamefont
  {Zhang}}\ and\ \bibinfo {author} {\bibfnamefont {A.}~\bibnamefont
  {Dalgarno}},\ }\href {\doibase 10.1080/00268970802126608} {\bibfield
  {journal} {\bibinfo  {journal} {Mol. Phys.}\ }\textbf {\bibinfo {volume}
  {106}},\ \bibinfo {pages} {1525} (\bibinfo {year} {2008})}\BibitemShut
  {NoStop}%
\bibitem [{\citenamefont {Salumbides}\ \emph {et~al.}(2014)\citenamefont
  {Salumbides}, \citenamefont {Ubachs},\ and\ \citenamefont
  {Korobov}}]{Salumbides2014}%
  \BibitemOpen
  \bibfield  {author} {\bibinfo {author} {\bibfnamefont {E.~J.}\ \bibnamefont
  {Salumbides}}, \bibinfo {author} {\bibfnamefont {W.}~\bibnamefont {Ubachs}},
  \ and\ \bibinfo {author} {\bibfnamefont {V.~I.}\ \bibnamefont {Korobov}},\
  }\href {\doibase 10.1016/j.jms.2014.04.003} {\bibfield  {journal} {\bibinfo
  {journal} {J. Mol. Spectrosc.}\ }\textbf {\bibinfo {volume} {300}},\ \bibinfo
  {pages} {65} (\bibinfo {year} {2014})}\BibitemShut {NoStop}%
\bibitem [{\citenamefont {Borkowski}\ \emph {et~al.}()\citenamefont
  {Borkowski}, \citenamefont {Buchachenko}, \citenamefont {Ciury{\l}o},
  \citenamefont {Julienne}, \citenamefont {Yamada}, \citenamefont {Kikuchi},
  \citenamefont {Takahashi}, \citenamefont {Takasu},\ and\ \citenamefont
  {Takahashi}}]{BorkowskiGravityTBP}%
  \BibitemOpen
  \bibfield  {author} {\bibinfo {author} {\bibfnamefont {M.}~\bibnamefont
  {Borkowski}}, \bibinfo {author} {\bibfnamefont {A.~A.}\ \bibnamefont
  {Buchachenko}}, \bibinfo {author} {\bibfnamefont {R.}~\bibnamefont
  {Ciury{\l}o}}, \bibinfo {author} {\bibfnamefont {P.~S.}\ \bibnamefont
  {Julienne}}, \bibinfo {author} {\bibfnamefont {H.}~\bibnamefont {Yamada}},
  \bibinfo {author} {\bibfnamefont {Y.}~\bibnamefont {Kikuchi}}, \bibinfo
  {author} {\bibfnamefont {K.}~\bibnamefont {Takahashi}}, \bibinfo {author}
  {\bibfnamefont {Y.}~\bibnamefont {Takasu}}, \ and\ \bibinfo {author}
  {\bibfnamefont {Y.}~\bibnamefont {Takahashi}},\ }\href@noop {} {\ ,\ \bibinfo
  {pages} {(to be published)}}\BibitemShut {NoStop}%
\bibitem [{\citenamefont {Takasu}\ \emph {et~al.}(2012)\citenamefont {Takasu},
  \citenamefont {Saito}, \citenamefont {Takahashi}, \citenamefont {Borkowski},
  \citenamefont {Ciury{\l}o},\ and\ \citenamefont {Julienne}}]{Takasu2012}%
  \BibitemOpen
  \bibfield  {author} {\bibinfo {author} {\bibfnamefont {Y.}~\bibnamefont
  {Takasu}}, \bibinfo {author} {\bibfnamefont {Y.}~\bibnamefont {Saito}},
  \bibinfo {author} {\bibfnamefont {Y.}~\bibnamefont {Takahashi}}, \bibinfo
  {author} {\bibfnamefont {M.}~\bibnamefont {Borkowski}}, \bibinfo {author}
  {\bibfnamefont {R.}~\bibnamefont {Ciury{\l}o}}, \ and\ \bibinfo {author}
  {\bibfnamefont {P.~S.}\ \bibnamefont {Julienne}},\ }\href {\doibase
  10.1103/PhysRevLett.108.173002} {\bibfield  {journal} {\bibinfo  {journal}
  {Phys. Rev. Lett.}\ }\textbf {\bibinfo {volume} {108}},\ \bibinfo {pages}
  {173002} (\bibinfo {year} {2012})}\BibitemShut {NoStop}%
\bibitem [{\citenamefont {McGuyer}\ \emph
  {et~al.}(2015{\natexlab{a}})\citenamefont {McGuyer}, \citenamefont
  {McDonald}, \citenamefont {Iwata}, \citenamefont {Tarallo}, \citenamefont
  {Skomorowski}, \citenamefont {Moszynski},\ and\ \citenamefont
  {Zelevinsky}}]{McGuyer2015}%
  \BibitemOpen
  \bibfield  {author} {\bibinfo {author} {\bibfnamefont {B.~H.}\ \bibnamefont
  {McGuyer}}, \bibinfo {author} {\bibfnamefont {M.}~\bibnamefont {McDonald}},
  \bibinfo {author} {\bibfnamefont {G.~Z.}\ \bibnamefont {Iwata}}, \bibinfo
  {author} {\bibfnamefont {M.~G.}\ \bibnamefont {Tarallo}}, \bibinfo {author}
  {\bibfnamefont {W.}~\bibnamefont {Skomorowski}}, \bibinfo {author}
  {\bibfnamefont {R.}~\bibnamefont {Moszynski}}, \ and\ \bibinfo {author}
  {\bibfnamefont {T.}~\bibnamefont {Zelevinsky}},\ }\href {\doibase
  10.1038/nphys3182} {\bibfield  {journal} {\bibinfo  {journal} {Nat. Phys.}\
  }\textbf {\bibinfo {volume} {11}},\ \bibinfo {pages} {32} (\bibinfo {year}
  {2015}{\natexlab{a}})}\BibitemShut {NoStop}%
\bibitem [{\citenamefont {McGuyer}\ \emph
  {et~al.}(2015{\natexlab{b}})\citenamefont {McGuyer}, \citenamefont
  {McDonald}, \citenamefont {Iwata}, \citenamefont {Tarallo}, \citenamefont
  {Grier}, \citenamefont {Apfelbeck},\ and\ \citenamefont
  {Zelevinsky}}]{McGuyer2015a}%
  \BibitemOpen
  \bibfield  {author} {\bibinfo {author} {\bibfnamefont {B.~H.}\ \bibnamefont
  {McGuyer}}, \bibinfo {author} {\bibfnamefont {M.}~\bibnamefont {McDonald}},
  \bibinfo {author} {\bibfnamefont {G.~Z.}\ \bibnamefont {Iwata}}, \bibinfo
  {author} {\bibfnamefont {M.~G.}\ \bibnamefont {Tarallo}}, \bibinfo {author}
  {\bibfnamefont {A.~T.}\ \bibnamefont {Grier}}, \bibinfo {author}
  {\bibfnamefont {F.}~\bibnamefont {Apfelbeck}}, \ and\ \bibinfo {author}
  {\bibfnamefont {T.}~\bibnamefont {Zelevinsky}},\ }\href {\doibase
  10.1088/1367-2630/17/5/055004} {\bibfield  {journal} {\bibinfo  {journal}
  {New J. Phys}\ }\textbf {\bibinfo {volume} {17}},\ \bibinfo {pages} {055004}
  (\bibinfo {year} {2015}{\natexlab{b}})}\BibitemShut {NoStop}%
\bibitem [{\citenamefont {McGuyer}\ \emph
  {et~al.}(2015{\natexlab{c}})\citenamefont {McGuyer}, \citenamefont
  {McDonald}, \citenamefont {Iwata}, \citenamefont {Skomorowski}, \citenamefont
  {Moszynski},\ and\ \citenamefont {Zelevinsky}}]{McGuyer2015b}%
  \BibitemOpen
  \bibfield  {author} {\bibinfo {author} {\bibfnamefont {B.~H.}\ \bibnamefont
  {McGuyer}}, \bibinfo {author} {\bibfnamefont {M.}~\bibnamefont {McDonald}},
  \bibinfo {author} {\bibfnamefont {G.~Z.}\ \bibnamefont {Iwata}}, \bibinfo
  {author} {\bibfnamefont {W.}~\bibnamefont {Skomorowski}}, \bibinfo {author}
  {\bibfnamefont {R.}~\bibnamefont {Moszynski}}, \ and\ \bibinfo {author}
  {\bibfnamefont {T.}~\bibnamefont {Zelevinsky}},\ }\href {\doibase
  10.1103/PhysRevLett.115.053001} {\bibfield  {journal} {\bibinfo  {journal}
  {Phys. Rev. Lett.}\ }\textbf {\bibinfo {volume} {115}},\ \bibinfo {pages}
  {053001} (\bibinfo {year} {2015}{\natexlab{c}})},\ \Eprint
  {http://arxiv.org/abs/1503.05946} {arXiv:1503.05946} \BibitemShut {NoStop}%
\bibitem [{\citenamefont {Taie}\ \emph {et~al.}(2016)\citenamefont {Taie},
  \citenamefont {Watanabe}, \citenamefont {Ichinose},\ and\ \citenamefont
  {Takahashi}}]{Taie2016}%
  \BibitemOpen
  \bibfield  {author} {\bibinfo {author} {\bibfnamefont {S.}~\bibnamefont
  {Taie}}, \bibinfo {author} {\bibfnamefont {S.}~\bibnamefont {Watanabe}},
  \bibinfo {author} {\bibfnamefont {T.}~\bibnamefont {Ichinose}}, \ and\
  \bibinfo {author} {\bibfnamefont {Y.}~\bibnamefont {Takahashi}},\ }\href
  {https://journals.aps.org/prl/pdf/10.1103/PhysRevLett.116.043202} {\bibfield
  {journal} {\bibinfo  {journal} {Phys. Rev. Lett.}\ }\textbf {\bibinfo
  {volume} {116}},\ \bibinfo {pages} {043202} (\bibinfo {year}
  {2016})}\BibitemShut {NoStop}%
\bibitem [{\citenamefont {Borkowski}\ \emph {et~al.}(2009)\citenamefont
  {Borkowski}, \citenamefont {Ciury{\l}o}, \citenamefont {Julienne},
  \citenamefont {Tojo}, \citenamefont {Enomoto},\ and\ \citenamefont
  {Takahashi}}]{Borkowski2009}%
  \BibitemOpen
  \bibfield  {author} {\bibinfo {author} {\bibfnamefont {M.}~\bibnamefont
  {Borkowski}}, \bibinfo {author} {\bibfnamefont {R.}~\bibnamefont
  {Ciury{\l}o}}, \bibinfo {author} {\bibfnamefont {P.~S.}\ \bibnamefont
  {Julienne}}, \bibinfo {author} {\bibfnamefont {S.}~\bibnamefont {Tojo}},
  \bibinfo {author} {\bibfnamefont {K.}~\bibnamefont {Enomoto}}, \ and\
  \bibinfo {author} {\bibfnamefont {Y.}~\bibnamefont {Takahashi}},\ }\href
  {\doibase 10.1103/PhysRevA.80.012715} {\bibfield  {journal} {\bibinfo
  {journal} {Phys. Rev. A}\ }\textbf {\bibinfo {volume} {80}},\ \bibinfo
  {pages} {012715} (\bibinfo {year} {2009})}\BibitemShut {NoStop}%
\bibitem [{\citenamefont {Taichenachev}\ \emph {et~al.}(2006)\citenamefont
  {Taichenachev}, \citenamefont {Yudin}, \citenamefont {Oates}, \citenamefont
  {Hoyt}, \citenamefont {Barber},\ and\ \citenamefont
  {Hollberg}}]{Taichenachev2006}%
  \BibitemOpen
  \bibfield  {author} {\bibinfo {author} {\bibfnamefont {A.~V.}\ \bibnamefont
  {Taichenachev}}, \bibinfo {author} {\bibfnamefont {V.~I.}\ \bibnamefont
  {Yudin}}, \bibinfo {author} {\bibfnamefont {C.~W.}\ \bibnamefont {Oates}},
  \bibinfo {author} {\bibfnamefont {C.~W.}\ \bibnamefont {Hoyt}}, \bibinfo
  {author} {\bibfnamefont {Z.~W.}\ \bibnamefont {Barber}}, \ and\ \bibinfo
  {author} {\bibfnamefont {L.}~\bibnamefont {Hollberg}},\ }\href {\doibase
  10.1103/PhysRevLett.96.083001} {\bibfield  {journal} {\bibinfo  {journal}
  {Phys. Rev. Lett.}\ }\textbf {\bibinfo {volume} {96}},\ \bibinfo {pages}
  {083001} (\bibinfo {year} {2006})}\BibitemShut {NoStop}%
\bibitem [{\citenamefont {Barber}\ \emph {et~al.}(2006)\citenamefont {Barber},
  \citenamefont {Hoyt}, \citenamefont {Oates}, \citenamefont {Hollberg},
  \citenamefont {Taichenachev},\ and\ \citenamefont {Yudin}}]{Barber2006}%
  \BibitemOpen
  \bibfield  {author} {\bibinfo {author} {\bibfnamefont {Z.~W.}\ \bibnamefont
  {Barber}}, \bibinfo {author} {\bibfnamefont {C.~W.}\ \bibnamefont {Hoyt}},
  \bibinfo {author} {\bibfnamefont {C.~W.}\ \bibnamefont {Oates}}, \bibinfo
  {author} {\bibfnamefont {L.}~\bibnamefont {Hollberg}}, \bibinfo {author}
  {\bibfnamefont {A.~V.}\ \bibnamefont {Taichenachev}}, \ and\ \bibinfo
  {author} {\bibfnamefont {V.~I.}\ \bibnamefont {Yudin}},\ }\href {\doibase
  10.1103/PhysRevLett.96.083002} {\bibfield  {journal} {\bibinfo  {journal}
  {Phys. Rev. Lett.}\ }\textbf {\bibinfo {volume} {96}},\ \bibinfo {pages}
  {083002} (\bibinfo {year} {2006})}\BibitemShut {NoStop}%
\bibitem [{\citenamefont {Franchi}\ \emph {et~al.}(2017)\citenamefont
  {Franchi}, \citenamefont {Livi}, \citenamefont {Cappellini}, \citenamefont
  {Binella}, \citenamefont {Inguscio}, \citenamefont {Catani},\ and\
  \citenamefont {Fallani}}]{Franchi2017}%
  \BibitemOpen
  \bibfield  {author} {\bibinfo {author} {\bibfnamefont {L.}~\bibnamefont
  {Franchi}}, \bibinfo {author} {\bibfnamefont {L.~F.}\ \bibnamefont {Livi}},
  \bibinfo {author} {\bibfnamefont {G.}~\bibnamefont {Cappellini}}, \bibinfo
  {author} {\bibfnamefont {G.}~\bibnamefont {Binella}}, \bibinfo {author}
  {\bibfnamefont {M.}~\bibnamefont {Inguscio}}, \bibinfo {author}
  {\bibfnamefont {J.}~\bibnamefont {Catani}}, \ and\ \bibinfo {author}
  {\bibfnamefont {L.}~\bibnamefont {Fallani}},\ }\href
  {http://iopscience.iop.org/article/10.1088/1367-2630/aa8fb4/pdf} {\bibfield
  {journal} {\bibinfo  {journal} {New J. Phys}\ }\textbf {\bibinfo {volume}
  {19}},\ \bibinfo {pages} {103037} (\bibinfo {year} {2017})}\BibitemShut
  {NoStop}%
\bibitem [{\citenamefont {Borkowski}\ \emph {et~al.}(2017)\citenamefont
  {Borkowski}, \citenamefont {Buchachenko}, \citenamefont {Ciury{\l}o},
  \citenamefont {Julienne}, \citenamefont {Yamada}, \citenamefont {Kikuchi},
  \citenamefont {Takahashi}, \citenamefont {Takasu},\ and\ \citenamefont
  {Takahashi}}]{Borkowski2017a}%
  \BibitemOpen
  \bibfield  {author} {\bibinfo {author} {\bibfnamefont {M.}~\bibnamefont
  {Borkowski}}, \bibinfo {author} {\bibfnamefont {A.~A.}\ \bibnamefont
  {Buchachenko}}, \bibinfo {author} {\bibfnamefont {R.}~\bibnamefont
  {Ciury{\l}o}}, \bibinfo {author} {\bibfnamefont {P.~S.}\ \bibnamefont
  {Julienne}}, \bibinfo {author} {\bibfnamefont {H.}~\bibnamefont {Yamada}},
  \bibinfo {author} {\bibfnamefont {Y.}~\bibnamefont {Kikuchi}}, \bibinfo
  {author} {\bibfnamefont {K.}~\bibnamefont {Takahashi}}, \bibinfo {author}
  {\bibfnamefont {Y.}~\bibnamefont {Takasu}}, \ and\ \bibinfo {author}
  {\bibfnamefont {Y.}~\bibnamefont {Takahashi}},\ }\href {\doibase
  10.1103/PhysRevA.96.063405} {\bibfield  {journal} {\bibinfo  {journal} {Phys.
  Rev. A}\ }\textbf {\bibinfo {volume} {96}},\ \bibinfo {pages} {063405}
  (\bibinfo {year} {2017})}\BibitemShut {NoStop}%
\bibitem [{\citenamefont {Bergmann}\ \emph {et~al.}(1998)\citenamefont
  {Bergmann}, \citenamefont {Theuer},\ and\ \citenamefont
  {Shore}}]{Bergmann1998}%
  \BibitemOpen
  \bibfield  {author} {\bibinfo {author} {\bibfnamefont {K.}~\bibnamefont
  {Bergmann}}, \bibinfo {author} {\bibfnamefont {H.}~\bibnamefont {Theuer}}, \
  and\ \bibinfo {author} {\bibfnamefont {B.~W.}\ \bibnamefont {Shore}},\ }\href
  {\doibase 10.1103/RevModPhys.70.1003} {\bibfield  {journal} {\bibinfo
  {journal} {Rev. Mod. Phys.}\ }\textbf {\bibinfo {volume} {70}},\ \bibinfo
  {pages} {1003} (\bibinfo {year} {1998})}\BibitemShut {NoStop}%
\bibitem [{\citenamefont {Stellmer}\ \emph {et~al.}(2012)\citenamefont
  {Stellmer}, \citenamefont {Pasquiou}, \citenamefont {Grimm},\ and\
  \citenamefont {Schreck}}]{Stellmer2012}%
  \BibitemOpen
  \bibfield  {author} {\bibinfo {author} {\bibfnamefont {S.}~\bibnamefont
  {Stellmer}}, \bibinfo {author} {\bibfnamefont {B.}~\bibnamefont {Pasquiou}},
  \bibinfo {author} {\bibfnamefont {R.}~\bibnamefont {Grimm}}, \ and\ \bibinfo
  {author} {\bibfnamefont {F.}~\bibnamefont {Schreck}},\ }\href {\doibase
  10.1103/PhysRevLett.109.115302} {\bibfield  {journal} {\bibinfo  {journal}
  {Phys. Rev. Lett.}\ }\textbf {\bibinfo {volume} {109}},\ \bibinfo {pages}
  {115302} (\bibinfo {year} {2012})}\BibitemShut {NoStop}%
\bibitem [{\citenamefont {Ciamei}\ \emph {et~al.}(2017)\citenamefont {Ciamei},
  \citenamefont {Bayerle}, \citenamefont {Chen}, \citenamefont {Pasquiou},\
  and\ \citenamefont {Schreck}}]{Ciamei2017}%
  \BibitemOpen
  \bibfield  {author} {\bibinfo {author} {\bibfnamefont {A.}~\bibnamefont
  {Ciamei}}, \bibinfo {author} {\bibfnamefont {A.}~\bibnamefont {Bayerle}},
  \bibinfo {author} {\bibfnamefont {C.-C.}\ \bibnamefont {Chen}}, \bibinfo
  {author} {\bibfnamefont {B.}~\bibnamefont {Pasquiou}}, \ and\ \bibinfo
  {author} {\bibfnamefont {F.}~\bibnamefont {Schreck}},\ }\href {\doibase
  10.1103/PhysRevA.96.013406} {\bibfield  {journal} {\bibinfo  {journal} {Phys.
  Rev. A}\ }\textbf {\bibinfo {volume} {96}},\ \bibinfo {pages} {013406}
  (\bibinfo {year} {2017})}\BibitemShut {NoStop}%
\bibitem [{\citenamefont {Tojo}\ \emph {et~al.}(2006)\citenamefont {Tojo},
  \citenamefont {Kitagawa}, \citenamefont {Enomoto}, \citenamefont {Kato},
  \citenamefont {Takasu}, \citenamefont {Kumakura},\ and\ \citenamefont
  {Takahashi}}]{Tojo2006}%
  \BibitemOpen
  \bibfield  {author} {\bibinfo {author} {\bibfnamefont {S.}~\bibnamefont
  {Tojo}}, \bibinfo {author} {\bibfnamefont {M.}~\bibnamefont {Kitagawa}},
  \bibinfo {author} {\bibfnamefont {K.}~\bibnamefont {Enomoto}}, \bibinfo
  {author} {\bibfnamefont {Y.}~\bibnamefont {Kato}}, \bibinfo {author}
  {\bibfnamefont {Y.}~\bibnamefont {Takasu}}, \bibinfo {author} {\bibfnamefont
  {M.}~\bibnamefont {Kumakura}}, \ and\ \bibinfo {author} {\bibfnamefont
  {Y.}~\bibnamefont {Takahashi}},\ }\href {\doibase
  10.1103/PhysRevLett.96.153201} {\bibfield  {journal} {\bibinfo  {journal}
  {Phys. Rev. Lett.}\ }\textbf {\bibinfo {volume} {96}},\ \bibinfo {pages}
  {153201} (\bibinfo {year} {2006})}\BibitemShut {NoStop}%
\bibitem [{\citenamefont {Bouganne}\ \emph {et~al.}(2017)\citenamefont
  {Bouganne}, \citenamefont {Aguilera}, \citenamefont {Dareau}, \citenamefont
  {Soave}, \citenamefont {Beugnon},\ and\ \citenamefont
  {Gerbier}}]{Bouganne2017}%
  \BibitemOpen
  \bibfield  {author} {\bibinfo {author} {\bibfnamefont {R.}~\bibnamefont
  {Bouganne}}, \bibinfo {author} {\bibfnamefont {M.~B.}\ \bibnamefont
  {Aguilera}}, \bibinfo {author} {\bibfnamefont {A.}~\bibnamefont {Dareau}},
  \bibinfo {author} {\bibfnamefont {E.}~\bibnamefont {Soave}}, \bibinfo
  {author} {\bibfnamefont {J.}~\bibnamefont {Beugnon}}, \ and\ \bibinfo
  {author} {\bibfnamefont {F.}~\bibnamefont {Gerbier}},\ }\href
  {https://doi.org/10.1088/1367-2630/aa8c45 http://arxiv.org/abs/1707.04307}
  {\bibfield  {journal} {\bibinfo  {journal} {New J. Phys.}\ }\textbf {\bibinfo
  {volume} {19}},\ \bibinfo {pages} {113006} (\bibinfo {year}
  {2017})}\BibitemShut {NoStop}%
\bibitem [{\citenamefont {Porsev}\ \emph {et~al.}(2014)\citenamefont {Porsev},
  \citenamefont {Safronova}, \citenamefont {Derevianko},\ and\ \citenamefont
  {Clark}}]{Porsev2014}%
  \BibitemOpen
  \bibfield  {author} {\bibinfo {author} {\bibfnamefont {S.~G.}\ \bibnamefont
  {Porsev}}, \bibinfo {author} {\bibfnamefont {M.~S.}\ \bibnamefont
  {Safronova}}, \bibinfo {author} {\bibfnamefont {A.}~\bibnamefont
  {Derevianko}}, \ and\ \bibinfo {author} {\bibfnamefont {C.~W.}\ \bibnamefont
  {Clark}},\ }\href {\doibase 10.1103/PhysRevA.89.012711} {\bibfield  {journal}
  {\bibinfo  {journal} {Phys. Rev. A}\ }\textbf {\bibinfo {volume} {89}},\
  \bibinfo {pages} {012711} (\bibinfo {year} {2014})}\BibitemShut {NoStop}%
\bibitem [{\citenamefont {Mies}\ \emph {et~al.}(1978)\citenamefont {Mies},
  \citenamefont {Stevens},\ and\ \citenamefont {Krauss}}]{Mies1978}%
  \BibitemOpen
  \bibfield  {author} {\bibinfo {author} {\bibfnamefont {F.~H.}\ \bibnamefont
  {Mies}}, \bibinfo {author} {\bibfnamefont {W.~J.}\ \bibnamefont {Stevens}}, \
  and\ \bibinfo {author} {\bibfnamefont {M.}~\bibnamefont {Krauss}},\ }\href
  {http://dacemirror.sci-hub.io/journal-article/e41dd6310bdac7384aecc02efe6c435a/mies1978.pdf}
  {\bibfield  {journal} {\bibinfo  {journal} {J. Mol. Spectrosc.}\ }\textbf
  {\bibinfo {volume} {72}},\ \bibinfo {pages} {303} (\bibinfo {year}
  {1978})}\BibitemShut {NoStop}%
\bibitem [{\citenamefont {Mohr}\ \emph {et~al.}(2016)\citenamefont {Mohr},
  \citenamefont {Taylor},\ and\ \citenamefont {Newell}}]{Mohr2016}%
  \BibitemOpen
  \bibfield  {author} {\bibinfo {author} {\bibfnamefont {P.}~\bibnamefont
  {Mohr}}, \bibinfo {author} {\bibfnamefont {B.}~\bibnamefont {Taylor}}, \ and\
  \bibinfo {author} {\bibfnamefont {D.}~\bibnamefont {Newell}},\ }\href
  {\doibase 10.1103/RevModPhys.84.1527} {\bibfield  {journal} {\bibinfo
  {journal} {Rev. Mod. Phys.}\ }\textbf {\bibinfo {volume} {88}},\ \bibinfo
  {pages} {035009} (\bibinfo {year} {2016})},\ \Eprint
  {http://arxiv.org/abs/0801.0028} {arXiv:0801.0028 [physics.atom-ph]}
  \BibitemShut {NoStop}%
\bibitem [{\citenamefont {Kitagawa}\ \emph {et~al.}(2008)\citenamefont
  {Kitagawa}, \citenamefont {Enomoto}, \citenamefont {Kasa}, \citenamefont
  {Takahashi}, \citenamefont {Ciury{\l}o}, \citenamefont {Naidon},\ and\
  \citenamefont {Julienne}}]{Kitagawa2008}%
  \BibitemOpen
  \bibfield  {author} {\bibinfo {author} {\bibfnamefont {M.}~\bibnamefont
  {Kitagawa}}, \bibinfo {author} {\bibfnamefont {K.}~\bibnamefont {Enomoto}},
  \bibinfo {author} {\bibfnamefont {K.}~\bibnamefont {Kasa}}, \bibinfo {author}
  {\bibfnamefont {Y.}~\bibnamefont {Takahashi}}, \bibinfo {author}
  {\bibfnamefont {R.}~\bibnamefont {Ciury{\l}o}}, \bibinfo {author}
  {\bibfnamefont {P.}~\bibnamefont {Naidon}}, \ and\ \bibinfo {author}
  {\bibfnamefont {P.~S.}\ \bibnamefont {Julienne}},\ }\href {\doibase
  10.1103/PhysRevA.77.012719} {\bibfield  {journal} {\bibinfo  {journal} {Phys.
  Rev. A}\ }\textbf {\bibinfo {volume} {77}},\ \bibinfo {pages} {012719}
  (\bibinfo {year} {2008})}\BibitemShut {NoStop}%
\bibitem [{\citenamefont {Atkins}\ and\ \citenamefont
  {de~Paula}(2010)}]{Atkins2010}%
  \BibitemOpen
  \bibfield  {author} {\bibinfo {author} {\bibfnamefont {P.}~\bibnamefont
  {Atkins}}\ and\ \bibinfo {author} {\bibfnamefont {J.}~\bibnamefont
  {de~Paula}},\ }\href@noop {} {\emph {\bibinfo {title} {{Atkins' Physical
  Chemistry}}}},\ \bibinfo {edition} {ninth edit}\ ed.\ (\bibinfo  {publisher}
  {Oxford University Press},\ \bibinfo {address} {Oxford},\ \bibinfo {year}
  {2010})\BibitemShut {NoStop}%
\bibitem [{\citenamefont {{Le Roy}}\ and\ \citenamefont
  {Bernstein}(1970)}]{Leroy1970}%
  \BibitemOpen
  \bibfield  {author} {\bibinfo {author} {\bibfnamefont {R.~J.}\ \bibnamefont
  {{Le Roy}}}\ and\ \bibinfo {author} {\bibfnamefont {R.~B.}\ \bibnamefont
  {Bernstein}},\ }\href {\doibase 10.1063/1.1697142} {\bibfield  {journal}
  {\bibinfo  {journal} {J. Chem. Phys.}\ }\textbf {\bibinfo {volume} {52}},\
  \bibinfo {pages} {3869} (\bibinfo {year} {1970})}\BibitemShut {NoStop}%
\bibitem [{\citenamefont {Gribakin}\ and\ \citenamefont
  {Flambaum}(1993)}]{Gribakin1993}%
  \BibitemOpen
  \bibfield  {author} {\bibinfo {author} {\bibfnamefont {G.~F.}\ \bibnamefont
  {Gribakin}}\ and\ \bibinfo {author} {\bibfnamefont {V.~V.}\ \bibnamefont
  {Flambaum}},\ }\href {\doibase 10.1103/PhysRevA.48.546} {\bibfield  {journal}
  {\bibinfo  {journal} {Phys. Rev. A}\ }\textbf {\bibinfo {volume} {48}},\
  \bibinfo {pages} {546} (\bibinfo {year} {1993})}\BibitemShut {NoStop}%
\bibitem [{\citenamefont {Ciury{\l}o}\ \emph {et~al.}(2004)\citenamefont
  {Ciury{\l}o}, \citenamefont {Tiesinga}, \citenamefont {Kotochigova},\ and\
  \citenamefont {Julienne}}]{Ciurylo2004}%
  \BibitemOpen
  \bibfield  {author} {\bibinfo {author} {\bibfnamefont {R.}~\bibnamefont
  {Ciury{\l}o}}, \bibinfo {author} {\bibfnamefont {E.}~\bibnamefont
  {Tiesinga}}, \bibinfo {author} {\bibfnamefont {S.}~\bibnamefont
  {Kotochigova}}, \ and\ \bibinfo {author} {\bibfnamefont {P.}~\bibnamefont
  {Julienne}},\ }\href {\doibase 10.1103/PhysRevA.70.062710} {\bibfield
  {journal} {\bibinfo  {journal} {Phys. Rev. A}\ }\textbf {\bibinfo {volume}
  {70}},\ \bibinfo {pages} {062710} (\bibinfo {year} {2004})}\BibitemShut
  {NoStop}%
\bibitem [{\citenamefont {Jones}\ \emph {et~al.}(2006)\citenamefont {Jones},
  \citenamefont {Tiesinga}, \citenamefont {Lett},\ and\ \citenamefont
  {Julienne}}]{Jones2006}%
  \BibitemOpen
  \bibfield  {author} {\bibinfo {author} {\bibfnamefont {K.~M.}\ \bibnamefont
  {Jones}}, \bibinfo {author} {\bibfnamefont {E.}~\bibnamefont {Tiesinga}},
  \bibinfo {author} {\bibfnamefont {P.~D.}\ \bibnamefont {Lett}}, \ and\
  \bibinfo {author} {\bibfnamefont {P.~S.}\ \bibnamefont {Julienne}},\ }\href
  {\doibase 10.1103/RevModPhys.78.483} {\bibfield  {journal} {\bibinfo
  {journal} {Rev. Mod. Phys.}\ }\textbf {\bibinfo {volume} {78}},\ \bibinfo
  {pages} {483} (\bibinfo {year} {2006})}\BibitemShut {NoStop}%
\bibitem [{\citenamefont {Lett}\ \emph {et~al.}(1993)\citenamefont {Lett},
  \citenamefont {Helmerson}, \citenamefont {Phillips}, \citenamefont {Ratliff},
  \citenamefont {Rolston},\ and\ \citenamefont {Wagshul}}]{Lett1993}%
  \BibitemOpen
  \bibfield  {author} {\bibinfo {author} {\bibfnamefont {P.~D.}\ \bibnamefont
  {Lett}}, \bibinfo {author} {\bibfnamefont {K.}~\bibnamefont {Helmerson}},
  \bibinfo {author} {\bibfnamefont {W.~D.}\ \bibnamefont {Phillips}}, \bibinfo
  {author} {\bibfnamefont {L.~P.}\ \bibnamefont {Ratliff}}, \bibinfo {author}
  {\bibfnamefont {S.~L.}\ \bibnamefont {Rolston}}, \ and\ \bibinfo {author}
  {\bibfnamefont {M.~E.}\ \bibnamefont {Wagshul}},\ }\href {\doibase
  10.1103/PhysRevLett.71.2200} {\bibfield  {journal} {\bibinfo  {journal}
  {Phys. Rev. Lett.}\ }\textbf {\bibinfo {volume} {71}},\ \bibinfo {pages}
  {2200} (\bibinfo {year} {1993})}\BibitemShut {NoStop}%
\bibitem [{\citenamefont {Abraham}\ \emph {et~al.}(1995)\citenamefont
  {Abraham}, \citenamefont {McAlexander}, \citenamefont {Sackett},\ and\
  \citenamefont {Hulet}}]{Abraham1995}%
  \BibitemOpen
  \bibfield  {author} {\bibinfo {author} {\bibfnamefont {E.~R.~I.}\
  \bibnamefont {Abraham}}, \bibinfo {author} {\bibfnamefont {W.~I.}\
  \bibnamefont {McAlexander}}, \bibinfo {author} {\bibfnamefont {C.~A.}\
  \bibnamefont {Sackett}}, \ and\ \bibinfo {author} {\bibfnamefont {R.~G.}\
  \bibnamefont {Hulet}},\ }\href {\doibase 10.1103/PhysRevLett.74.1315}
  {\bibfield  {journal} {\bibinfo  {journal} {Phys. Rev. Lett.}\ }\textbf
  {\bibinfo {volume} {74}},\ \bibinfo {pages} {1315} (\bibinfo {year}
  {1995})}\BibitemShut {NoStop}%
\bibitem [{\citenamefont {Tsai}\ \emph {et~al.}(1997)\citenamefont {Tsai},
  \citenamefont {Freeland}, \citenamefont {Vogels}, \citenamefont {Boesten},
  \citenamefont {Verhaar},\ and\ \citenamefont {Heinzen}}]{Tsai1997}%
  \BibitemOpen
  \bibfield  {author} {\bibinfo {author} {\bibfnamefont {C.~C.}\ \bibnamefont
  {Tsai}}, \bibinfo {author} {\bibfnamefont {R.~S.}\ \bibnamefont {Freeland}},
  \bibinfo {author} {\bibfnamefont {J.~M.}\ \bibnamefont {Vogels}}, \bibinfo
  {author} {\bibfnamefont {H.~M. J.~M.}\ \bibnamefont {Boesten}}, \bibinfo
  {author} {\bibfnamefont {B.~J.}\ \bibnamefont {Verhaar}}, \ and\ \bibinfo
  {author} {\bibfnamefont {D.~J.}\ \bibnamefont {Heinzen}},\ }\href {\doibase
  10.1103/PhysRevLett.79.1245} {\bibfield  {journal} {\bibinfo  {journal}
  {Phys. Rev. Lett.}\ }\textbf {\bibinfo {volume} {79}},\ \bibinfo {pages}
  {1245} (\bibinfo {year} {1997})}\BibitemShut {NoStop}%
\bibitem [{\citenamefont {{Martinez De Escobar}}\ \emph
  {et~al.}(2008)\citenamefont {{Martinez De Escobar}}, \citenamefont
  {Mickelson}, \citenamefont {Pellegrini}, \citenamefont {Nagel}, \citenamefont
  {Traverso}, \citenamefont {Yan}, \citenamefont {C{\^{o}}t{\'{e}}},\ and\
  \citenamefont {Killian}}]{MartinezDeEscobar2008}%
  \BibitemOpen
  \bibfield  {author} {\bibinfo {author} {\bibfnamefont {Y.~N.}\ \bibnamefont
  {{Martinez De Escobar}}}, \bibinfo {author} {\bibfnamefont {P.~G.}\
  \bibnamefont {Mickelson}}, \bibinfo {author} {\bibfnamefont {P.}~\bibnamefont
  {Pellegrini}}, \bibinfo {author} {\bibfnamefont {S.~B.}\ \bibnamefont
  {Nagel}}, \bibinfo {author} {\bibfnamefont {A.}~\bibnamefont {Traverso}},
  \bibinfo {author} {\bibfnamefont {M.}~\bibnamefont {Yan}}, \bibinfo {author}
  {\bibfnamefont {R.}~\bibnamefont {C{\^{o}}t{\'{e}}}}, \ and\ \bibinfo
  {author} {\bibfnamefont {T.~C.}\ \bibnamefont {Killian}},\ }\href {\doibase
  10.1103/PhysRevA.78.062708} {\bibfield  {journal} {\bibinfo  {journal} {Phys.
  Rev. A}\ }\textbf {\bibinfo {volume} {78}},\ \bibinfo {pages} {062708}
  (\bibinfo {year} {2008})}\BibitemShut {NoStop}%
\bibitem [{\citenamefont {M{\"{u}}nchow}\ \emph {et~al.}(2011)\citenamefont
  {M{\"{u}}nchow}, \citenamefont {Bruni}, \citenamefont {Madalinski},\ and\
  \citenamefont {G{\"{o}}rlitz}}]{Munchow2011}%
  \BibitemOpen
  \bibfield  {author} {\bibinfo {author} {\bibfnamefont {F.}~\bibnamefont
  {M{\"{u}}nchow}}, \bibinfo {author} {\bibfnamefont {C.}~\bibnamefont
  {Bruni}}, \bibinfo {author} {\bibfnamefont {M.}~\bibnamefont {Madalinski}}, \
  and\ \bibinfo {author} {\bibfnamefont {A.}~\bibnamefont {G{\"{o}}rlitz}},\
  }\href {\doibase 10.1039/c1cp21219b} {\bibfield  {journal} {\bibinfo
  {journal} {Phys. Chem. Chem. Phys.}\ }\textbf {\bibinfo {volume} {13}},\
  \bibinfo {pages} {18734} (\bibinfo {year} {2011})}\BibitemShut {NoStop}%
\bibitem [{\citenamefont {Kahmann}\ \emph {et~al.}(2014)\citenamefont
  {Kahmann}, \citenamefont {Tiemann}, \citenamefont {Appel}, \citenamefont
  {Sterr},\ and\ \citenamefont {Riehle}}]{Kahmann2014}%
  \BibitemOpen
  \bibfield  {author} {\bibinfo {author} {\bibfnamefont {M.}~\bibnamefont
  {Kahmann}}, \bibinfo {author} {\bibfnamefont {E.}~\bibnamefont {Tiemann}},
  \bibinfo {author} {\bibfnamefont {O.}~\bibnamefont {Appel}}, \bibinfo
  {author} {\bibfnamefont {U.}~\bibnamefont {Sterr}}, \ and\ \bibinfo {author}
  {\bibfnamefont {F.}~\bibnamefont {Riehle}},\ }\href {\doibase
  10.1103/PhysRevA.89.023413} {\bibfield  {journal} {\bibinfo  {journal} {Phys.
  Rev. A}\ }\textbf {\bibinfo {volume} {89}},\ \bibinfo {pages} {023413}
  (\bibinfo {year} {2014})}\BibitemShut {NoStop}%
\bibitem [{\citenamefont {Colbert}\ \emph {et~al.}(1992)\citenamefont
  {Colbert}, \citenamefont {Miller}, \citenamefont {Colbert},\ and\
  \citenamefont {Miller}}]{Colbert1992}%
  \BibitemOpen
  \bibfield  {author} {\bibinfo {author} {\bibfnamefont {D.~T.}\ \bibnamefont
  {Colbert}}, \bibinfo {author} {\bibfnamefont {W.~H.}\ \bibnamefont {Miller}},
  \bibinfo {author} {\bibfnamefont {D.~T.}\ \bibnamefont {Colbert}}, \ and\
  \bibinfo {author} {\bibfnamefont {W.~H.}\ \bibnamefont {Miller}},\ }\href
  {\doibase 10.1063/1.462100} {\bibfield  {journal} {\bibinfo  {journal} {J.
  Chem. Phys.}\ }\textbf {\bibinfo {volume} {96}},\ \bibinfo {pages} {1982}
  (\bibinfo {year} {1992})}\BibitemShut {NoStop}%
\bibitem [{\citenamefont {Tiesinga}\ \emph {et~al.}(1998)\citenamefont
  {Tiesinga}, \citenamefont {Williams},\ and\ \citenamefont
  {Julienne}}]{Tiesinga1998}%
  \BibitemOpen
  \bibfield  {author} {\bibinfo {author} {\bibfnamefont {E.}~\bibnamefont
  {Tiesinga}}, \bibinfo {author} {\bibfnamefont {C.~J.}\ \bibnamefont
  {Williams}}, \ and\ \bibinfo {author} {\bibfnamefont {P.~S.}\ \bibnamefont
  {Julienne}},\ }\href {\doibase 10.1103/PhysRevA.57.4257} {\bibfield
  {journal} {\bibinfo  {journal} {Phys. Rev. A}\ }\textbf {\bibinfo {volume}
  {57}},\ \bibinfo {pages} {4257} (\bibinfo {year} {1998})}\BibitemShut
  {NoStop}%
\bibitem [{\citenamefont {Angeli}\ and\ \citenamefont
  {Marinova}(2013)}]{Angeli2011}%
  \BibitemOpen
  \bibfield  {author} {\bibinfo {author} {\bibfnamefont {I.}~\bibnamefont
  {Angeli}}\ and\ \bibinfo {author} {\bibfnamefont {K.~P.}\ \bibnamefont
  {Marinova}},\ }\href {\doibase 10.1016/j.adt.2011.12.006} {\bibfield
  {journal} {\bibinfo  {journal} {At. Data Nucl. Data Tables}\ }\textbf
  {\bibinfo {volume} {99}},\ \bibinfo {pages} {69} (\bibinfo {year}
  {2013})}\BibitemShut {NoStop}%
\bibitem [{\citenamefont {Gao}(1996)}]{Gao1996}%
  \BibitemOpen
  \bibfield  {author} {\bibinfo {author} {\bibfnamefont {B.}~\bibnamefont
  {Gao}},\ }\href {\doibase 10.1103/PhysRevA.54.2022} {\bibfield  {journal}
  {\bibinfo  {journal} {Phys. Rev. A}\ }\textbf {\bibinfo {volume} {54}},\
  \bibinfo {pages} {2022} (\bibinfo {year} {1996})}\BibitemShut {NoStop}%
\bibitem [{\citenamefont {Power}(1967)}]{Power1967}%
  \BibitemOpen
  \bibfield  {author} {\bibinfo {author} {\bibfnamefont {E.~A.}\ \bibnamefont
  {Power}},\ }\href {\doibase 10.1063/1.1840541} {\bibfield  {journal}
  {\bibinfo  {journal} {J. Chem. Phys.}\ }\textbf {\bibinfo {volume} {46}},\
  \bibinfo {pages} {4297} (\bibinfo {year} {1967})}\BibitemShut {NoStop}%
\bibitem [{\citenamefont {Vitanov}\ and\ \citenamefont
  {Stenholm}(1997)}]{Vitanov1997}%
  \BibitemOpen
  \bibfield  {author} {\bibinfo {author} {\bibfnamefont {N.~V.}\ \bibnamefont
  {Vitanov}}\ and\ \bibinfo {author} {\bibfnamefont {S.}~\bibnamefont
  {Stenholm}},\ }\href {\doibase 10.1103/PhysRevA.56.1463} {\bibfield
  {journal} {\bibinfo  {journal} {Phys. Rev. A}\ }\textbf {\bibinfo {volume}
  {56}},\ \bibinfo {pages} {1463} (\bibinfo {year} {1997})}\BibitemShut
  {NoStop}%
\bibitem [{\citenamefont {Winkler}\ \emph {et~al.}(2007)\citenamefont
  {Winkler}, \citenamefont {Lang}, \citenamefont {Thalhammer}, \citenamefont
  {van~der Straten}, \citenamefont {Grimm},\ and\ \citenamefont
  {Denschlag}}]{Winkler2007}%
  \BibitemOpen
  \bibfield  {author} {\bibinfo {author} {\bibfnamefont {K.}~\bibnamefont
  {Winkler}}, \bibinfo {author} {\bibfnamefont {F.}~\bibnamefont {Lang}},
  \bibinfo {author} {\bibfnamefont {G.}~\bibnamefont {Thalhammer}}, \bibinfo
  {author} {\bibfnamefont {P.}~\bibnamefont {van~der Straten}}, \bibinfo
  {author} {\bibfnamefont {R.}~\bibnamefont {Grimm}}, \ and\ \bibinfo {author}
  {\bibfnamefont {J.~H.}\ \bibnamefont {Denschlag}},\ }\href {\doibase
  10.1103/PhysRevLett.98.043201} {\bibfield  {journal} {\bibinfo  {journal}
  {Phys. Rev. Lett.}\ }\textbf {\bibinfo {volume} {98}},\ \bibinfo {pages}
  {043201} (\bibinfo {year} {2007})},\ \Eprint {http://arxiv.org/abs/0611222}
  {arXiv:0611222 [cond-mat]} \BibitemShut {NoStop}%
\bibitem [{\citenamefont {Fioretti}\ \emph {et~al.}(1998)\citenamefont
  {Fioretti}, \citenamefont {Comparat}, \citenamefont {Crubellier},
  \citenamefont {Dulieu}, \citenamefont {Masnou-Seeuws},\ and\ \citenamefont
  {Pillet}}]{Fioretti1998}%
  \BibitemOpen
  \bibfield  {author} {\bibinfo {author} {\bibfnamefont {A.}~\bibnamefont
  {Fioretti}}, \bibinfo {author} {\bibfnamefont {D.}~\bibnamefont {Comparat}},
  \bibinfo {author} {\bibfnamefont {A.}~\bibnamefont {Crubellier}}, \bibinfo
  {author} {\bibfnamefont {O.}~\bibnamefont {Dulieu}}, \bibinfo {author}
  {\bibfnamefont {F.}~\bibnamefont {Masnou-Seeuws}}, \ and\ \bibinfo {author}
  {\bibfnamefont {P.}~\bibnamefont {Pillet}},\ }\href {\doibase
  10.1103/PhysRevLett.80.4402} {\bibfield  {journal} {\bibinfo  {journal}
  {Phys. Rev. Lett.}\ }\textbf {\bibinfo {volume} {80}},\ \bibinfo {pages}
  {4402} (\bibinfo {year} {1998})}\BibitemShut {NoStop}%
\bibitem [{\citenamefont {Reinaudi}\ \emph {et~al.}(2012)\citenamefont
  {Reinaudi}, \citenamefont {Osborn}, \citenamefont {McDonald}, \citenamefont
  {Kotochigova},\ and\ \citenamefont {Zelevinsky}}]{Reinaudi2012}%
  \BibitemOpen
  \bibfield  {author} {\bibinfo {author} {\bibfnamefont {G.}~\bibnamefont
  {Reinaudi}}, \bibinfo {author} {\bibfnamefont {C.~B.}\ \bibnamefont
  {Osborn}}, \bibinfo {author} {\bibfnamefont {M.}~\bibnamefont {McDonald}},
  \bibinfo {author} {\bibfnamefont {S.}~\bibnamefont {Kotochigova}}, \ and\
  \bibinfo {author} {\bibfnamefont {T.}~\bibnamefont {Zelevinsky}},\ }\href
  {\doibase 10.1103/PhysRevLett.109.115303} {\bibfield  {journal} {\bibinfo
  {journal} {Phys. Rev. Lett.}\ }\textbf {\bibinfo {volume} {109}},\ \bibinfo
  {pages} {115303} (\bibinfo {year} {2012})}\BibitemShut {NoStop}%
\end{thebibliography}%
\end{document}